\documentclass{aa}  

\usepackage{graphicx}
\usepackage{txfonts}
\usepackage{lipsum}
\usepackage{natbib}
\usepackage{xspace}
\usepackage[normalem]{ulem}

\usepackage{hyperref}
\usepackage{subcaption}         
\usepackage{lscape}             
\usepackage{placeins}           
                                
\usepackage[table,xcdraw]{xcolor}
\definecolor{krishnacomment}{HTML}{ff0000}

\begin{document}

   \title{$\nu$Gundam: a unified framework for point-redshift and PDF-aware spatial correlation functions}

   \author{E. Donoso\inst{1,2}
          \and
          C. A. Lopez Cortez\inst{1,2}
          \and
          U. Sureshkumar\inst{3, 4}
          \and
          Mariano J. Domínguez Romero \inst{5, 6}
          }

   \institute{
        Instituto de Ciencias Astronómicas, de la Tierra y del Espacio (ICATE), Av. Nazario Benavidez 8175 Oeste, Chimbas, San Juan, Argentina; Universidad Nacional de San Juan, Argentina; Consejo Nacional de Investigaciones Científicas y Técnicas (CONICET), Argentina.
        \email{emiliodon@gmail.com}
        \and
        Facultad de Ciencias Exactas, Físicas y Naturales (FCEFYN-UNSJ), Universidad Nacional de San Juan, Av. José Ignacio de la Roza Oeste 590, San Juan, Argentina.
        \and
        National Centre for Nuclear Research, ul. Pasteura 7, 02-093 Warsaw, Poland.
        \and
        Wits Centre for Astrophysics, School of Physics, University of the Witwatersrand, Private Bag 3, Johannesburg 2050, South Africa
        \and
        Instituto de Astronom\'{\i}a Te\'orica y Experimental (IATE -  UNC and CONICET CCT C\'ordoba), \\Observatorio Astron\'omico de C\'ordoba, Universidad Nacional de C\'ordoba, Laprida 854, X5000BGR, C\'ordoba, Argentina.
        \and
        Centro de Investigaciones de Frontera, Universidad de La Serena, Ra\'ul Bitr\'an 1305, La Serena, Chile.
        }

    \abstract
    {Photometric-redshift uncertainties redistribute galaxy pairs along the line of sight (LOS) and across projected-separation bins, diluting spatial clustering measurements. Modern photo-$z$ methods provide full per-object redshift probability density functions (PDFs), but propagating them directly through standard pair-count estimators remains computationally demanding.}
    {We present \scalebox{1.2}{$\nu$}\textsc{Gundam}\xspace, a unified framework for point-redshift and PDF-aware spatial correlation functions. Our aim is to propagate radial PDFs at the pair-count level while retaining a common interface for exact, compressed, and stochastic projected-clustering calculations.}
    {The code implements four PDF-aware modes: exact empirical-PDF integration on a shared radial grid, Gaussian-mixture (GMM) compression, Monte Carlo (MC) sampling, and a quantile-compression method 16quant. All modes accumulate pair counts on a resolved $(r_\mathrm{p},\pi)$ grid, from which $\xi(r_\mathrm{p},\pi)$ and cumulative $w_\mathrm{p}(r_\mathrm{p};\Pi)$ are reconstructed. We validate these methods with DESI galaxies and CosmoDC2 mocks using pair-level diagnostics, controlled degradation, convergence tests, and an exact-grid benchmark.}
    {The PDF-aware modes produce nearly indistinguishable LOS redistribution kernels, while point-photo-$z$ estimates discard probabilistic pair information. Controlled Gaussian and empirical-PDF experiments show that radial uncertainty mainly redistributes existing clustering signal to larger $\pi$, and that the LOS depth required to recover the truth-redshift projection scales approximately with the effective pairwise radial dispersion. For practical settings, MC, 16quant, and GMM reproduce the exact-grid result with fractional root-mean-square (RMS) residuals below $5\%$. Applied to DESI red and blue Bright Galaxy Survey samples, MC sampling recovers their spectroscopic clustering shapes and relative amplitudes for $r_\mathrm{p}\gtrsim0.8\,h^{-1}\mathrm{Mpc}$, using sample-dependent recovery depths of approximately $400$ and $180\,h^{-1}\mathrm{Mpc}$, respectively.}
    {Projected clustering with redshift PDFs is best understood as a pair-redistribution problem. When propagated consistently, the projected signal can be recovered accurately and efficiently. The LOS integration depth is nevertheless sample dependent and should be treated as part of the measurement definition rather than as a fixed technical choice.}


   \maketitle
   \nolinenumbers

\section{Introduction}
Two-point correlation functions remain one of the central tools for deriving cosmological information from galaxy surveys. In spectroscopic datasets, precise redshift measurements allow mapping the three-dimensional clustering of galaxies, whether in redshift space or through projected correlations, across a wide range of scales. Imaging surveys, however, deliver much larger and deeper galaxy samples at the price of substantially less precise radial information. In practice, the line-of-sight (LOS) uncertainty associated with photometric redshifts dilutes the clustering signal, mixes widely separated pairs, and can bias inferences if it is handled only through a single-point redshift estimate or through overly broad tomographic bins. At the same time, modern photo-$z$ methods, based on template fitting, machine learning, or hybrid approaches, increasingly provide a full per-object redshift probability density function (PDF), often with non-Gaussian structure and multiple modes, rather than only a mean, mode, or median estimate \citep{salvato2019,schmidt2020}.

This has led to several distinct strategies for incorporating photometric-redshift uncertainty into clustering analyses. One well-established approach is to work with angular or tomographic clustering, and to deproject the signal statistically using the redshift distribution of the sample. In that spirit, \citet{arnaltemur2009} showed with mock photometric surveys that the real-space correlation function \citep{zhou2021_desi} can be recovered by deprojection when the photo-$z$ uncertainty is modeled appropriately. A second route is to use spectroscopic objects as radial anchors and exploit the full PDF of the photometric sample in a cross-correlation estimator. \citet{myers2009} introduced a real-space spectroscopic--photometric estimator that weights pairs by the photometric PDF and showed substantial gains in signal-to-noise relative to estimators based only on the PDF peak. Related work has applied PDF-weighted projected and redshift-space clustering estimators to galaxy and active galactic nuclei (AGN) samples containing both spectroscopic and photometric redshifts, demonstrating the practical value of retaining full redshift distributions in spatial clustering analyses \citep{georgakakis2014,allevato2016}. A third route is to retain the clustering analysis in projection or in tomographic slices but replace hard sample assignment by PDF-weighted sample membership. Using mock DES-like catalogs, \citet{asorey2016} compared mean-, mode-, and full-PDF-based binning schemes and found that using the whole PDF reduces the bias in angular clustering and galaxy-bias measurements, particularly for narrow redshift bins.

\citet{zhou2021_desi} adopted a complementary strategy for the clustering of photometrically selected luminous red galaxies. They cross-correlated galaxies in a target photo-$z$ interval
with a second sample extending beyond that interval by the adopted LOS integration depth, thereby reducing the loss of eligible pairs across redshift-bin boundaries. Photometric-redshift uncertainties were incorporated through forward modeling by perturbing the LOS positions of galaxies in the mock catalogs according to the observed error distribution. This approach addresses boundary leakage and radial smearing at the sample-selection and modeling levels, whereas the methods developed here marginalize the per-object radial PDFs directly
within the observed pair counts, including skewed and multimodal uncertainties.

PDF-aware pair counting is relevant across a broad range of scales. On large scales, realistic redshift PDFs affect the modeling of projected and transverse baryon acoustic oscillation (BAO) measurements \citep{chavesmontero2018,chan2022,ferreira2026}. On small scales, full PDFs have been used to assign probabilistic close-pair weights and infer merger fractions in photometric samples \citep{lopezsanjuan2015,mundy2017,duncan2019}, with recent extensions to AGN pairs and JWST samples \citep{dougherty2024,duan2025}. Both applications require propagating the joint radial uncertainty of each pair through a geometrical separation criterion, which is the central operation implemented by \scalebox{1.2}{$\nu$}\textsc{Gundam}\xspace.

These studies make clear that full redshift PDFs carry useful clustering information and that, in many regimes, they outperform point-estimate-based treatments. They also expose a practical gap. Existing approaches are often tailored to a specific observable or geometry: angular clustering with tomographic bins, spectroscopic--photometric cross-correlations with a fixed anchored sample, specialized projected estimators developed for particular galaxy or AGN samples, or theory-level modeling of particular 3D photometric statistics. By contrast, there has been comparatively little emphasis on a general-purpose, pair-count-based framework that can incorporate full per-object PDFs directly into standard spatial correlation-function estimators, while still remaining fast enough for modern survey catalogs. This is particularly relevant for projected clustering measurements, where one would ideally propagate the radial uncertainty of each pair into the line-of-sight integration, rather than collapse every galaxy to a single redshift before counting.

The computational difficulty is straightforward: once each object is represented by a PDF, a pair no longer contributes to a single separation bin, but to a distribution over line-of-sight and, in general, projected separations. An exact treatment based on empirical PDFs sampled on a common grid is attractive because it preserves arbitrary shapes and multimodality, but it can be expensive in both memory and runtime. This naturally motivates compressed or stochastic approximations that are fast enough for production use but still faithful to the exact PDF-aware calculation. Establishing when these approximations succeed, and what aspects of the clustering signal they preserve or distort, is therefore a central methodological problem.

In this paper, we present \scalebox{1.2}{$\nu$}\textsc{Gundam}\xspace, a new version of the original \textsc{Gundam} toolkit \citep{donoso2019}, designed to compute spatial correlation functions in the presence of photometric-redshift uncertainty. The new framework implements four complementary PDF-aware modes. The first is an exact empirical-PDF mode, in which pair weights are evaluated deterministically from per-object PDFs defined on a common radial grid. The second is a Gaussian-mixture (GMM) mode, which compresses each empirical PDF into a small number of moment-matched Gaussian components and evaluates line-of-sight shell probabilities analytically in comoving-distance space. The third is a Monte Carlo (MC) mode, in which standard point-redshift pair counts are averaged over repeated realizations drawn from the input PDFs. The fourth is a quantile-compression mode, 16quant, in which each empirical PDF is represented by a fixed set of equal-probability quantile nodes whose Cartesian products are propagated through the same $(r_\mathrm{p},\pi)$ pair-counting geometry. These modes are implemented within a common multi-$\pi$ pair-counting framework, so that exact, compressed, and stochastic treatments can be compared on the same footing and applied with the same analysis interface.

We validate and illustrate these methods using two complementary data sets. The first is an observed sample of DESI spectroscopic galaxies, for which we construct empirical photometric-redshift PDFs and point estimates from random-forest predictions based on photometry. DESI provides a large, precise spectroscopic reference sample for clustering studies and large-scale-structure applications \citep{DESI2016}. The second is a mock galaxy sample based on CosmoDC2, a synthetic sky catalog developed for Legacy Survey of Space and Time (LSST)/Rubin dark-energy analyses, which provides a controlled environment in which to test the behavior of the methods under survey conditions representative of future deep photometric data \citep{korytov2019}. Together, these two applications allow us to test the methods both on real observed galaxies and on realistic LSST-like simulations.

The goals of this paper are therefore twofold. First, we provide a methodological and software description of a practical framework for PDF-aware spatial correlation functions, including exact and approximate algorithms that can be run within the same code base. Second, we assess the fidelity of these methods with dedicated validation tests that isolate line-of-sight kernel recovery, quantify the degradation of projected clustering under controlled PDF broadening, and map the regimes in which exact-grid, GMM, and Monte Carlo approaches are most reliable. The paper is organized as follows. In Sect.~\ref{sec:software} we describe the underlying software architecture of \scalebox{1.2}{$\nu$}\textsc{Gundam}\xspace. In Sect.~\ref{sec:methods} we describe the formalism and its implementation in \scalebox{1.2}{$\nu$}\textsc{Gundam}\xspace. In Sect.~\ref{sec:validation} we introduce the validation framework, summarize the behavior of the four PDF-aware modes, and present applications with DESI survey data and CosmoDC2 mock samples. We summarize our conclusions in Sect.~\ref{sec:conclusions}.

\section{Underlying software framework}
\label{sec:software}

\scalebox{1.2}{$\nu$}\textsc{Gundam}\xspace\footnote{https://github.com/samotracio/nugundam} preserves the core architecture of the original \textsc{Gundam} package \citep{donoso2019}. Catalog handling, configuration, and I/O are managed in Python, whereas expensive pair counts are performed by compiled, highly optimized Fortran kernels that run in parallel via OpenMP. Acceleration relies on linked lists and cell-level skip lists built on a spatial grid, so that large regions of the catalog can be skipped when they cannot contribute to a given separation bin. This general strategy used for both angular and projected space is extremely effective in counting pairs with high throughput. 

Several public packages provide efficient pair-counting and large-scale-structure analysis for catalogs represented by scalar positions, including \textsc{CUTE}, \textsc{Corrfunc}, \textsc{nbodykit}, and \textsc{FCFC} \citep{alonso2012,sinha2020,hand2018,zhao2023}. The distinctive role of \scalebox{1.2}{$\nu$}\textsc{Gundam}\xspace is to make per-object radial probability distributions native inputs to the same projected counting, estimator, resampling, and output workflow.

The new code is built around a completely rewritten Python wrapper and presents some methodological extensions. First, in addition to bootstrap support, the package now provides completely automatic jackknife realizations and covariance estimates, with regions either supplied as input, or generated automatically from the survey footprint. For reference, the Landy--Szalay estimator \citep{landyszalay1993} used throughout this work is
\begin{equation}
\xi_{\rm LS} = \frac{DD - 2\,DR + RR}{RR},
\label{eq:ls_estimator_early}
\end{equation}
where DD, DR, and RR denote the normalized data--data, data--random, and random--random pair counts, respectively.
Second, for Landy--Szalay auto-correlations, \scalebox{1.2}{$\nu$}\textsc{Gundam}\xspace implements a split-random option inspired by \citet{keihanen2019}, in which the random catalog is divided into smaller chunks in order to reduce the cost of the RR term while preserving the estimator normalization. Third, the package includes marked correlation functions \citep[e.g.][]{skibba2006, sureshkumar2021}, so that environment-dependent clustering can be measured within the same configuration and result framework as the standard two-point statistics. These additions are integrated into the same preparation, counting, resampling, and plotting pipeline used by the PDF-aware estimators described below.

Internal jackknife and bootstrap covariance estimates are convenient for survey analyses, although their accuracy can be scale dependent; precision applications should therefore validate them against mock catalogs where possible \citep{norberg2009}.

\section{Methods}
\label{sec:methods}

\subsection{Projected clustering and notation}
\label{subsec:methods:projected-notation}

We consider the projected two-point correlation function of galaxy samples specified by angular positions on the sky and by comoving radial distances. For a galaxy pair with angular separation $\theta_{ij}$ and comoving radial distances $\chi_1$ and $\chi_2$, we define
\begin{equation}
    \pi \equiv |\chi_1-\chi_2|,
    \qquad
    r_\mathrm{p} \simeq \theta_{ij}\sqrt{\chi_1\chi_2} .
    \label{eq:rp_pi_def}
\end{equation}

The approximate sign in the definition of $r_\mathrm{p}$ is the result of the small-angle approximation. The exact angular contribution to the pair separation is 
$r_\mathrm{p}=2\sqrt{\chi_1\chi_2}\sin(\theta_{ij}/2)$, which reduces to 
$r_\mathrm{p}\simeq\theta_{ij}\sqrt{\chi_1\chi_2}$ for $\theta_{ij}\ll1$, with fractional corrections of order $\theta_{ij}^2$. In practice, $\theta_{ij}$ is evaluated from the
three-dimensional unit vectors associated with the two sky directions, so the
projected-separation test is performed through
$r_\mathrm{p}^2 \simeq \theta_{ij}^2\chi_1\chi_2$. The geometric mean
$\sqrt{\chi_1\chi_2}$ provides a symmetric effective radial distance for the
pair. For pairs with $\chi_1 \simeq \chi_2$, it is equivalent to the more
familiar midpoint approximation at leading order.

The projected correlation function is widely used to reduce sensitivity to redshift-space distortions by integrating $\xi(r_\mathrm{p},\pi)$ along the LOS \citep{davispeebles1983,zehavi2011}.

For point-estimate redshifts, the projected correlation function is obtained
from the two-dimensional redshift-space correlation function
$\xi(r_\mathrm{p},\pi)$ through
\begin{equation}
    w_\mathrm{p}(r_\mathrm{p}) = 2 \int_0^{\pi_{\max}} \xi(r_\mathrm{p},\pi)\,d\pi .
    \label{eq:wp_continuous}
\end{equation}
In the discretized implementation, the projected-separation bins are indexed by $b$ and the LOS bins by $v$. If the LOS grid has edges
$\{\pi_v\}_{v=0}^{N_\pi}$, with $\Delta\pi_v \equiv \pi_{v+1}-\pi_v$ and $\pi_{\max}=\pi_{N_\pi}$, then the projected correlation function in the $r_{\mathrm{p}, b}$ bin integrated up to the outer edge $\pi_{\max}$ is reconstructed as
\begin{equation}
    w_\mathrm{p}(r_{\mathrm{p},b};\pi_{\max})
    \simeq
    2\sum_{v=0}^{N_\pi-1}\xi_{bv}\,\Delta\pi_v .
    \label{eq:wp_discrete_final}
\end{equation}
The same two-dimensional count fields can also be used to construct cumulative
projected measurements as a function of LOS depth,
\begin{equation}
    w_\mathrm{p}(r_{\mathrm{p},b};\Pi_q)
    \simeq
    2\sum_{v=0}^{q-1}\xi_{bv}\,\Delta\pi_v,
    \qquad
    \Pi_q \equiv \pi_q,
    \qquad
    q=1,\ldots,N_\pi .
    \label{eq:wp_discrete_cumulative}
\end{equation}
Thus, the fundamental measured object in the projected calculation is the
binned field $\xi_{bv}\equiv\xi(r_{\mathrm{p},b},\pi_v)$, or equivalently the set of
DD, DR, and RR pair-count fields from which it is estimated. The projected
quantity $w_\mathrm{p}$ is obtained only after summing these LOS shells.

The classical point-estimate mode is recovered by assigning each galaxy a
single distance estimate $\hat{\chi}_i$ and counting each pair in the unique
cell of the $(r_\mathrm{p},\pi)$ grid determined by
$(\hat{\chi}_i,\hat{\chi}_j)$. In contrast, the PDF-aware functionality
introduced below replaces this hard assignment by an expected contribution to
one or more $(r_\mathrm{p},\pi)$ bins, obtained after marginalizing over the
redshift--distance uncertainty of the two galaxies. Throughout this section,
all PDFs are written in comoving-distance space, because the geometry entering
Eqs.~(\ref{eq:rp_pi_def})--(\ref{eq:wp_discrete_cumulative}) is naturally
expressed in $\chi$. When input PDFs are supplied on a common redshift grid, \scalebox{1.2}{$\nu$}\textsc{Gundam}\xspace converts the shared support to comoving distance
using the chosen cosmology before evaluating the projected pair weights.

The LOS binning deserves one additional comment. The case $N_\pi=1$
corresponds to a single, wide, projected window, with one bin spanning
$0\leq\pi<\pi_{\max}$. Although this option is useful for fast projected-window measurements, it should not be assumed to be numerically identical to a measurement performed with several narrower LOS bins sharing the same outer edge $\pi_{\max}$. In the resolved case, the estimator is evaluated on each $(r_\mathrm{p},\pi)$ shell and the projected statistic is then formed through Eq.~(\ref{eq:wp_discrete_final}), whereas in a single-wide-bin measurement the counts are first accumulated over the full LOS interval and the estimator is applied only to that coarse bin. Since the standard estimators are nonlinear ratios of DD, DR, and RR counts, these two discretizations can differ for a finite bin width. The resolved-$\pi$ calculation is therefore the natural discretized form of Eq.~(\ref{eq:wp_continuous}), while the single-bin case should be interpreted as a coarse projected-window approximation.

\subsection{A generic PDF-aware pair-weight formulation}
\label{subsec:methods:generic-pair-weight}

Let $p_i(\chi)$ denote the radial probability distribution of object $i$ in comoving-distance space. In the PDF-aware projected calculation, a pair is no longer assigned to a single location on the $(r_\mathrm{p},\pi)$ grid. Instead, its contribution is distributed over the bins allowed by the two radial PDFs. For a projected-separation bin $b$ and a line-of-sight bin $v$, we define the bin-indicator kernel
\begin{equation}
\Theta_{bv}(\theta_{ij},\chi_1,\chi_2)=
\begin{cases}
1, & r_{\mathrm{p},b} \leq \theta_{ij}\sqrt{\chi_1\chi_2} < r_{\mathrm{p},b+1}, \\
   & \pi_v \leq |\chi_1-\chi_2| < \pi_{v+1}, \\
0, & \mathrm{otherwise} .
\end{cases}
\label{eq:pdf_bin_indicator_2d}
\end{equation}
The expected contribution of pair $(i,j)$ to bin $(b,v)$ is then
\begin{equation}
W^{(bv)}_{ij}
=
\omega_i\omega_j
\int d\chi_1 \int d\chi_2\,
p_i(\chi_1)p_j(\chi_2)
\Theta_{bv}(\theta_{ij},\chi_1,\chi_2),
\label{eq:pdf_pair_weight_2d}
\end{equation}
where $\omega_i$ and $\omega_j$ are optional object-level weights. These weights describe sample or systematic weights and are distinct from the radial PDF weighting. In the spectroscopic limit, $p_i(\chi)=\delta_{\rm D}(\chi-\chi_i)$, Eq.~\eqref{eq:pdf_pair_weight_2d} reduces to the usual hard assignment of a point-redshift pair to one bin of the $(r_\mathrm{p},\pi)$ grid.

The corresponding expected pair-count fields are obtained by summing these pair weights over the relevant catalog pairs,
\begin{equation}
C_{bv} = \sum_{\langle i,j\rangle} W^{(bv)}_{ij},
\label{eq:pdf_expected_counts}
\end{equation}
with unordered pairs for auto-correlations and pairs drawn from the two input catalogs for cross-correlations. The same construction is used for the data--data (DD), data--random (DR), and random--random (RR) terms, with the appropriate PDFs and object weights for each catalog.

Equation~\eqref{eq:pdf_pair_weight_2d} is the common statistical target of the four PDF-aware modes implemented in \scalebox{1.2}{$\nu$}\textsc{Gundam}\xspace. The exact empirical-PDF mode evaluates the integral deterministically on the common PDF grid. The GMM mode replaces each empirical PDF by a small moment-matched Gaussian mixture and evaluates the line-of-sight shell probabilities analytically. The Monte Carlo mode draws scalar distances from the PDFs, performs standard point-redshift counts on each realization, and averages the resulting count fields. Once the expected DD, DR, and RR arrays have been accumulated on the $(r_\mathrm{p},\pi)$ grid, the estimator algebra is unchanged: the standard natural (NAT), Davis--Peebles (DP), or Landy--Szalay (LS) estimators are applied \citep{peebleshauser1974,davispeebles1983,landyszalay1993} to obtain $\xi_{bv}$, and $w_p(r_\mathrm{p})$ is reconstructed by the line-of-sight sum described in Sect.~\ref{subsec:methods:projected-notation}.

\subsection{Exact empirical-PDF integration on a common grid}
\label{subsec:methods:exact-grid}

In the exact empirical-PDF mode (referred to as exact or ePDF mode), each radial PDF is represented by a normalized probability vector on a shared comoving-distance grid $p_{im}\equiv p_i(\chi_m)$. For cross-correlations, the per-object PDFs of the two input catalogs are first represented on a common $\chi$ grid. The expected contribution of a pair to the two-dimensional bin $(b,v)$ is then the discrete form of Eq.~\eqref{eq:pdf_pair_weight_2d},
\begin{equation}
W^{(bv)}_{ij,{\rm ePDF}}
=
\omega_i\omega_j
\sum_m \sum_n
p_{im}p_{jn}\,
\Theta_{bv}(\theta_{ij},\chi_m,\chi_n) .
\label{eq:epdf_double_sum_2d}
\end{equation}
For a fixed input grid, this calculation produces the same pair weights on every run and introduces no additional approximation beyond the discrete representation of the empirical PDFs. It is therefore ideal to use it as the reference against which faster approximate methods can be compared: the Gaussian-mixture and quantile modes, which replace each PDF with a compact representation, and the Monte Carlo mode, which estimates the same pair weights by repeated random sampling.

A direct evaluation of Eq.~\eqref{eq:epdf_double_sum_2d} would be expensive. The implementation reduces the cost by using cumulative probability sums. For a given distance $\chi_m$ of object $i$, projected bin $b$, and cumulative LOS edge $\Pi$, the allowed values of  $\chi_n$ for object $j$ can be obtained using the condition $|\chi_m-\chi_n|<\Pi$. If $A_{j,bm}(\Pi)$ denotes the probability mass of PDF $j$ in this interval, evaluated from the cumulative distribution of $p_j$, the shell contribution is computed as
\begin{equation}
W^{(bv)}_{ij,{\rm ePDF}}
=
\omega_i\omega_j
\sum_m p_{im}
\left[
A_{j,bm}(\pi_{v+1})-A_{j,bm}(\pi_v)
\right] .
\label{eq:epdf_cumulative_difference}
\end{equation}
Thus the multi-$\pi$ calculation is obtained by differencing cumulative masses at the requested $\pi$-bin edges, rather than by performing an independent double sum for every LOS shell.

For input PDFs supplied as edge-grid histograms, \scalebox{1.2}{$\nu$}\textsc{Gundam}\xspace can refine each input bin into several equal-width sub-bins in $\chi$ and distribute the original bin probability uniformly among them. This edge refinement provides a controlled quadrature approximation to the continuous histogram model and reduces radial quantization artifacts when the requested $\pi$ bins are narrower than, or comparable to, the PDF-grid spacing. Setting the refinement factor to unity recovers the unrefined grid representation.

The exact-grid counters also use active-support pruning. For each PDF, only grid points above a configurable probability floor are stored in the sparse active support; if no grid point survives, the peak bin is retained. These support bounds conservatively increase the radial search depth and reject candidate pairs whose PDF supports cannot overlap within the outer LOS edge. This pruning changes only the computational cost, not the statistical target of Eq.~\eqref{eq:epdf_double_sum_2d}. Because the allowed projected bin depends explicitly on both $\chi_m$ and $\chi_n$, the exact empirical-PDF mode captures migration in both $\pi$ and $r_\mathrm{p}$ induced by the radial PDFs.

\subsection{Gaussian-mixture compression in comoving distance}
\label{subsec:methods:gmm}

The GMM mode provides a fast deterministic approximation to Eq.~\eqref{eq:pdf_pair_weight_2d}. Each empirical radial PDF is replaced by a small Gaussian mixture in comoving-distance space,
\begin{equation}
p_i(\chi) \simeq
\sum_{a=1}^{K}\alpha_{ia}\,
{\cal N}(\chi\mid \mu_{ia},\sigma_{ia}^2),
\qquad
\alpha_{ia}\geq 0,
\qquad
\sum_{a=1}^{K}\alpha_{ia}=1 .
\label{eq:gmm_mixture}
\end{equation}
This mixture should be interpreted as a deterministic compression of the empirical PDF, not as a Gaussian mixture fitted via expectation maximization. In the default compressor, the empirical cumulative distribution is divided into $K$ segments with approximately equal probability mass. For each segment, \scalebox{1.2}{$\nu$}\textsc{Gundam}\xspace stores the mass of the segment, conditional mean and conditional variance, with a small variance floor to avoid degenerate components. When input PDFs are supplied as edge-grid histograms, we treat the probability in each bin as uniformly distributed over the corresponding interval in $\chi$.
The segment boundaries are then allowed to fall inside these input bins rather than only at the bin edges.
We analytically account for the contribution of the finite bin width to the variance.
This edge-aware compression reduces artificial radial quantization from coarse PDF grids in the resulting GMM representation.

For a pair of mixture components $a$ and $c$, belonging to objects $i$ and $j$, the LOS separation is a Gaussian random variable $X=\chi_i-\chi_j$ with mean $m_{iacj}=\mu_{ia}-\mu_{jc}$ and variance $s_{iacj}^2=\sigma_{ia}^2+\sigma_{jc}^2$. The probability that this component pair contributes to the LOS shell $v$ is therefore
\begin{equation}
\begin{split}
P^{(v)}_{ia,cj}
&=
F_{|X|}\!\left(\pi_{v+1};m_{iacj},s_{iacj}\right)
-
F_{|X|}\!\left(\pi_{v};m_{iacj},s_{iacj}\right),
\end{split}
\label{eq:gmm_shell_probability}
\end{equation}
where $F_{|X|}(x;m,s)=
\Phi[(x-m)/s]-\Phi[(-x-m)/s]$ and $\Phi$ is the standard normal cumulative distribution function. The single-wide-window case is recovered by setting $\pi_0=0$ and $\pi_1=\pi_{\max}$.

Let $B_{b,ia,cj}(\theta_{ij})$ be equal to unity when
$r_{\mathrm{p},b} \leq \theta_{ij}\sqrt{\mu_{ia}\mu_{jc}} < r_{\mathrm{p},b+1}$
and zero otherwise. The compiled GMM counters then accumulate
the expected contribution
\begin{equation}
W^{(bv)}_{ij,\mathrm{GMM}}
=
\omega_i\omega_j
\sum_{a=1}^{K}
\sum_{c=1}^{K}
\alpha_{ia}\alpha_{jc}\,
B_{b,ia,cj}(\theta_{ij})\,
P^{(v)}_{ia,cj}.
\label{eq:gmm_pair_weight_2d}
\end{equation}
Thus, the LOS probability is distributed analytically over the requested $\pi$ shells, while the projected-separation bin is assigned from the component-pair means. This is the main approximation of the current GMM mode relative to the exact empirical-PDF calculation: radial uncertainty is propagated in the LOS direction, but migration across $r_\mathrm{p}$ bins is not integrated over the full two-dimensional $(\chi_1,\chi_2)$ support of the component pair.

The GMM counters are embedded in the same linked-cell search used by the point-redshift and exact-grid projected kernels. The radial search depth is conservatively enlarged using summary information from the PDF representation, and candidate component pairs with negligible mixture weight are skipped. These pruning steps affect the runtime but not the expression in Eq.~\eqref{eq:gmm_pair_weight_2d}. For fixed PDF assignments, the same shell-probability construction is also used by the compiled jackknife and bootstrap GMM fast counters, so the deterministic GMM calculation can be combined with the resampling machinery on both single-wide and resolved-$\pi$ grids.

\subsection{Monte Carlo PDF sampling}
\label{subsec:methods:mc}

The MC mode provides a stochastic approximation to the same PDF-marginalized count fields described by Eq.~\eqref{eq:pdf_pair_weight_2d}. For each realization $r=1,\ldots,R$, one scalar comoving distance is drawn for every object, $\chi_i^{(r)}\sim p_i(\chi)$, and the standard point-redshift projected counter is run on the resulting catalog. Since the ordinary projected backend already accumulates counts on the full $(r_\mathrm{p},\pi)$ grid, the MC mode naturally supports both single-wide and resolved-$\pi$ measurements.

For any count term $C$, such as DD, DR, or RR in the auto-correlation case, the realization-averaged count field is
\begin{equation}
\overline{C}_{bv}
=
{1\over R}
\sum_{r=1}^{R} C^{(r)}_{bv} .
\label{eq:mc_count_average}
\end{equation}
The chosen estimator is then applied to the averaged count fields $\overline{\rm DD}_{bv}$, $\overline{\rm DR}_{bv}$, and $\overline{\rm RR}_{bv}$, and $w_p(r_\mathrm{p})$ is obtained from the resulting $\xi_{bv}$ as in Eq.~\eqref{eq:wp_discrete_final}. Thus, \scalebox{1.2}{$\nu$}\textsc{Gundam}\xspace estimates the expectation of pair counts before reconstructing the correlation function. Averaging the counts, rather than the final $w_p$ curves, keeps the estimator algebra identical to the point-redshift pipeline and gives a direct stochastic counterpart of the deterministic PDF modes.

Because each realization assigns a full scalar distance to every object, the MC method samples migration in both $\pi$ and $r_\mathrm{p}$. In the limit of many realizations it converges to the PDF-marginalized count fields, with sampling fluctuations that decrease approximately as $R^{-1/2}$ at fixed catalog size and binning. Individual realization-level $w_p$ curves can be stored as a diagnostic of MC noise.

When the input PDFs are supplied as edge-grid histograms, the selected PDF bin can either be represented by its center or sampled continuously inside its bin edges. The latter option treats the probability in each input bin as uniformly distributed across the corresponding interval in $\chi$, and is useful when the requested $\pi$ bins are narrower than or comparable to the spacing of the PDF grid. This within-bin sampling is the MC analog of the edge-aware GMM compression and the edge-refined exact-grid calculation.

If explicit random-catalog PDFs are supplied, random distances are sampled from those PDFs in the same way as for the data. Otherwise, \scalebox{1.2}{$\nu$}\textsc{Gundam}\xspace assigns random distances from the data-PDF information using one of three policies. In the \texttt{fixed-global} policy, a single random radial realization is drawn from the global mean data PDF and reused across MC realizations, reducing noise in the random terms and keeping RR fixed. In the \texttt{rerun-global} policy, a fresh random radial realization is drawn from the same global mean PDF for each MC realization. In the \texttt{inherit-realization} policy, random distances are sampled with replacement from the current realization of the data distances, which closely couples the random radial selection to the realized data distribution. For cross-correlations, the same logic is applied independently to the two sides of the measurement.

\subsection{Quantile compression in comoving distance: 16quant}
\label{subsec:methods:quantile-chi}

The 16quant mode is a deterministic compressed-grid approximation to the same PDF-marginalized pair weight in Eq.~\eqref{eq:pdf_pair_weight_2d}. Instead of representing each empirical PDF by Gaussian components or by the full input grid, each object is represented by a fixed number $N_q$ of equal-probability quantile nodes in comoving distance (by default $N_q=16$).

Quantile representations have previously been investigated as compact formats for large photo-$z$ PDF catalogs, with their fidelity depending on both the properties of the PDF ensemble and the downstream science metric \citep{malz2018}. The 16quant mode differs in using the quantile nodes directly inside the pair counter, rather than reconstructing a continuous PDF before evaluating the observable.

Let $F_i(\chi)$ be the cumulative distribution of the radial PDF of object $i$, and let $Q_i(u)=F_i^{-1}(u)$ be its inverse CDF. The quantile nodes are placed at the midpoint probabilities
\begin{equation}
    u_a={a-1/2 \over N_q},
    \qquad
    \chi_{ia}=Q_i(u_a),
    \qquad
    a=1,\ldots,N_q .
    \label{eq:qchi_nodes}
\end{equation}
The empirical PDF is then approximated by an equal-weight discrete measure over these nodes. For edge-grid PDF inputs, the probability in each input bin is treated as uniformly distributed across the corresponding interval in $\chi$, so the inverse CDF is interpolated inside the bin. This is the quantile analogue of the edge-aware treatments used by the GMM, exact-grid, and MC modes, and reduces artificial radial quantization from coarse PDF grids.

The expected contribution of a pair to bin $(b,v)$ is accumulated as the Cartesian product of the two quantile sets,
\begin{equation}
    W^{(bv)}_{ij,{\rm Q}}
    =
    {\omega_i\omega_j \over N_q^2}
    \sum_{a=1}^{N_q}
    \sum_{c=1}^{N_q}
    \Theta_{bv}(\theta_{ij},\chi_{ia},\chi_{jc}) .
    \label{eq:qchi_pair_weight}
\end{equation}
Therefore, each quantile pair is tested with the same geometry as used by the projection kernels of the point-redshift and the exact-grid, namely $\pi=|\chi_{ia}-\chi_{jc}|$ and $r_\mathrm{p}^2 \simeq \theta_{ij}^2\chi_{ia}\chi_{jc}$. The method consequently propagates the radial uncertainty in both $\pi$ and $r_\mathrm{p}$, but only at the finite resolution set by $N_q$. Unlike MC, it has no realization noise, and unlike GMM, it makes no Gaussian assumption and does not assign the projected separation from a component-pair mean.

The compiled 16quant counters are embedded in the same linked-cell and skip-list search used by the other projected kernels. For each object, \scalebox{1.2}{$\nu$}\textsc{Gundam}\xspace stores the quantile library together with the mean quantile position and the lower and upper quantile limits. These summaries are used to sort the catalog, set conservative angular search distances, and prune candidate pairs whose quantile supports cannot overlap inside the requested LOS or projected-separation range. Only candidate pairs that pass these support tests enter the $N_q^2$ quantile-product loop.

The current implementation supports both single-wide and resolved multi-$\pi$ projected measurements, with weighted and unweighted counts for auto- and cross-correlations. Native fast bootstrap and jackknife-touch kernels are implemented for unweighted 16quant configurations. As for the other compiled deterministic PDF modes, random catalogs inherit quantile-PDF assignments from the associated data sample.

\subsection{Random catalogs in the compiled PDF modes}
\label{subsec:methods:randoms-compiled-pdf}

For the compiled deterministic PDF modes (GMM and exact-grid), the current implementation assumes that random catalogs do not carry their own PDF library.
Instead, randoms inherit PDF assignments from the associated data sample by drawing, with replacement, indices into the prepared data-PDF library.
This policy is used for auto-correlations and separately on each side of cross-correlations.
The inherited random distances and PDF indices are therefore consistent with the same compressed or exact PDF representation used for the data objects. This design choice keeps the memory footprint moderate and avoids storing a second large PDF library for the random catalogs.

\subsection{Choosing the LOS integration scale}
\label{subsec:methods:pimax}

A projected analysis still requires a choice of $\pi_{\max}$.
Previous projected-clustering analyses using photometric redshifts adopted substantially enlarged LOS windows and, in some cases, padded redshift selections to accommodate radial
smearing and pair loss across sample boundaries \citep{georgakakis2014,allevato2016,zhou2021_desi}. This motivates a sample-dependent prescription rather than the direct adoption of a fixed value calibrated for spectroscopic redshifts.
For PDF-aware clustering, this scale should be large enough to contain the effective pairwise radial smearing of the sample, but not so large that noise and unrelated foreground-background pairs dominate the signal.
\scalebox{1.2}{$\nu$}\textsc{Gundam}\xspace implements a method that estimates a sensible default directly from the empirical PDFs by first building an effective pairwise radial width $\sigma_{\rm pw,eff}$ from robust summaries of the variance distribution per-object. The exact expression depends on whether the measurement is an auto-correlation, a photo--photo cross-correlation, or a spec--photo cross-correlation.
The suggested LOS window is then
\begin{equation}
\pi_{\max}^{\rm guess}=m_\pi\,\sigma_{\rm pw,eff},
\label{eq:pimax-guess}
\end{equation}
with default multiplier $m_\pi=2.0$.
This prescription is not part of the estimator itself, but it provides a reproducible data-driven starting point for the projected integration depth used in the experiments below.

\begin{figure*}
    \centering
    \includegraphics[width=1.0\linewidth]{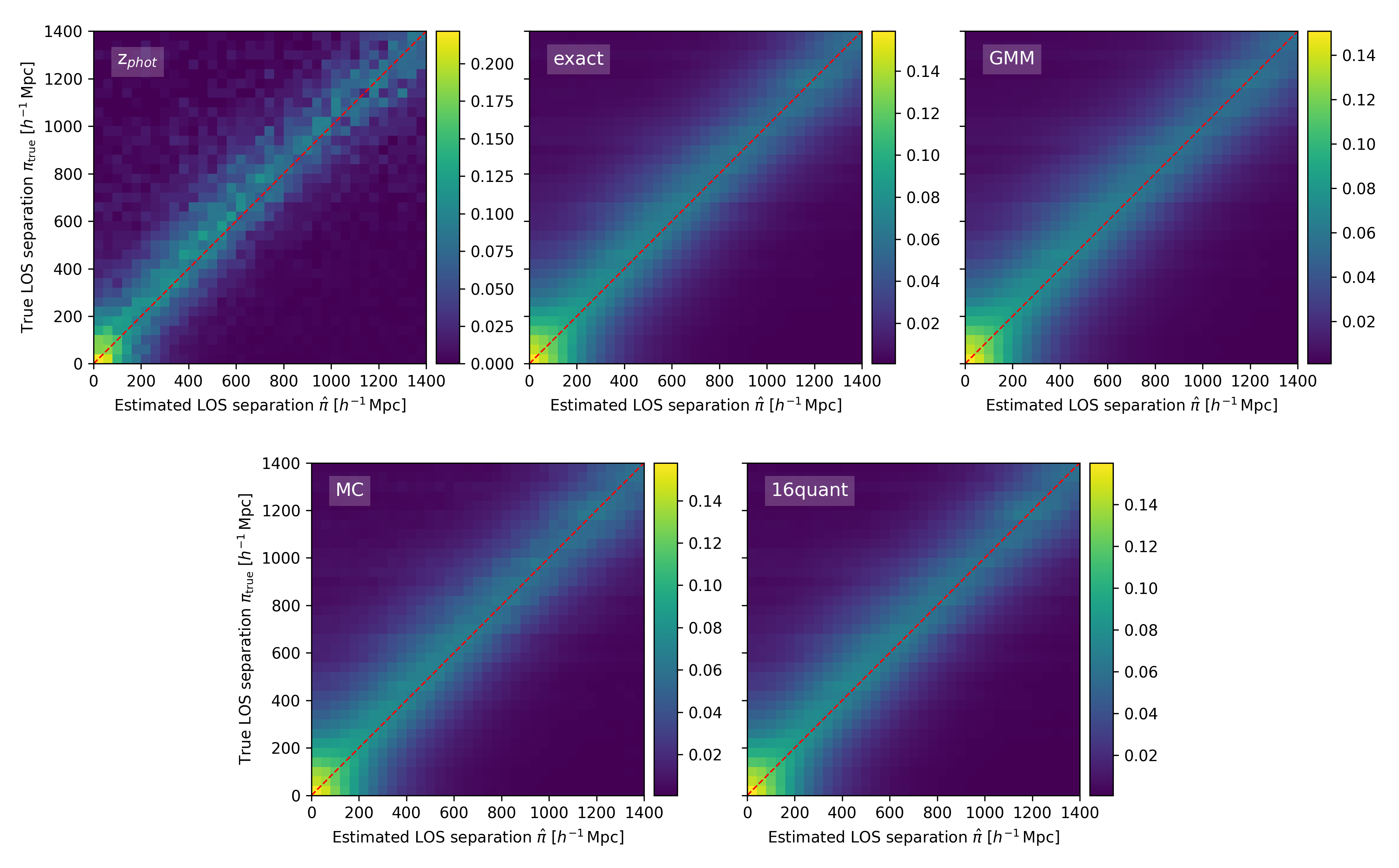}
    \caption{LOS pair-response for $30\,000$ randomly selected galaxy pairs from the DESI sample at $z<1.7$, restricted to a narrow, true projected separation range ($1-5~h^{-1}\rm Mpc$). Each panel shows the mean probability assigned to estimated LOS separation $\hat\pi$ at fixed true LOS separation $\pi_{\rm true}$. Colors indicate the mean pair response per $\hat\pi$ of each method. The PDF modes produce nearly identical smooth LOS response matrices, while the point-estimate ($z_{phot}$) case shows a more structured and concentrated response.
    }
    \label{fig:kernel_pair_response}
\end{figure*}

\subsection{Summary and practical interpretation of the PDF-aware modes}
\label{subsec:methods:summary}
The present PDF-aware implementation is targeted at projected clustering measurements, but the fundamental accumulated quantities are pair-count fields on the two-dimensional $(r_\mathrm{p},\pi)$ grid. The projected statistic $w_p(r_\mathrm{p})$ is then obtained by summing these fields over the chosen LOS range. The single-wide-bin case, with one bin spanning $0\leq\pi<\pi_{\max}$, can be a useful coarse projected-window mode, while resolved multi-$\pi$ binning allows the construction of $\xi(r_\mathrm{p},\pi)$ and cumulative projected measurements $w_p(r_\mathrm{p};\Pi)$.

The four PDF-aware modes differ in how they approximate the same PDF-marginalized pair-count problem, as listed below.

\begin{description}
    \item[\texttt{Exact-grid} mode:] The exact-grid mode evaluates Eq.~(\ref{eq:epdf_double_sum_2d}) deterministically on the shared empirical grid and captures pair migration in both $\pi$ and $r_\mathrm{p}$, but is the most computationally expensive.
    \item[\texttt{GMM} mode:] The GMM mode compresses each empirical PDF into a small number of Gaussian components, evaluates LOS shell probabilities analytically, and assigns projected separation from component-pair means.
    \item[\texttt{MC} mode:] The MC mode draws one scalar distance per galaxy and realization, reuses the standard projected counters, and converges to the PDF-marginalized result as the number of realizations increases.
    \item[\texttt{16quant} mode:] The 16quant mode replaces each empirical PDF by a small set of equal-probability quantile nodes and accumulates the deterministic Cartesian product of those nodes using the same $(r_\mathrm{p},\pi)$ geometry as the projected pair counter.
\end{description}

These differences define the trade-off space explored in the validation sections that follow. Exact-grid integration provides the deterministic reference where feasible. GMM is compact and smooth in the LOS direction, but its treatment of $r_\mathrm{p}$ migration is approximate. MC is simple and asymptotically faithful, but introduces realization noise. 16quant occupies an intermediate regime: it is deterministic and much cheaper than exact-grid integration, while retaining a discrete representation of the radial PDF that can move probability across both LOS and projected-separation bins. In particular, the pair-response diagnostic primarily probes LOS weighting at fixed true $r_\mathrm{p}$, while the degradation experiment measures the combined impact of LOS smearing, $r_\mathrm{p}$ migration, and estimator-level normalization on the final $w_p(r_\mathrm{p})$ estimate.

\section{Validation framework and diagnostic experiments}
\label{sec:validation}

In this section, we validate the PDF-aware estimators implemented in \scalebox{1.2}{$\nu$}\textsc{Gundam}\xspace and quantify the information gained by propagating redshift uncertainty at the pair-counting level.  We begin by examining the line-of-sight pair response of the different methods, which shows how point estimates, Monte Carlo sampling, Gaussian-mixture compression,
empirical PDFs, and quantile representations redistribute pair contributions as a function of $\Pi$.  We then introduce a complementary pair-probability calibration test, designed to isolate the information lost when a full redshift PDF is compressed into a hard point-photo-$z$
decision.  This diagnostic compares the PDF-predicted probability that a pair lies within a given $\Pi_{\max}$ to the actual fraction of such pairs that are close in true comoving distance.  Having established the pair-level behavior of the estimators, we proceed to controlled degradation tests using mock Gaussian and double-Gaussian redshift
uncertainties, and finally to measurements based on the full photometric
redshift PDFs of the CosmoDC2 galaxies.  Together, these tests separate
three questions: how redshift uncertainty broadens the line-of-sight response, whether full PDFs encode meaningful probabilistic pair information beyond scalar photo-$z$ estimates, and under what integration scale the recovered projected clustering approaches the $z_{\rm true}$
reference signal.

The four validation tests described below use two different galaxy samples, chosen according to what each test requires. The LOS pair-response diagnostic (Sect.~\ref{subsec:validation:pair-response}) uses real DESI spectroscopic data described in detail in Appendix ~\ref{sec:DESI}: because DESI provides secure spectroscopic redshifts for a large sample of real galaxies, the spec-z can be used directly as the "true" LOS separation against which each method's estimated hat(pi) is compared, without requiring a simulated catalog. The remaining three tests instead use the CosmoDC2 mock catalog, described in detail in Appendix ~\ref{sec:cosmodc2}. These tests require not only a known truth redshift for every galaxy, but also the ability to attach synthetic or realistic photometric-redshift PDFs of controlled width and shape to the same objects, and in some cases to compare against an exact empirical-PDF calculation. The estimation of photometric redshifts is described in detail in  Appendix ~\ref{sec:photoz}.
This combination of known truth and flexible, controllable PDF assignment is only available in a mock catalog, which is why CosmoDC2 rather than DESI is used for these three tests.

\subsection{LOS pair-response diagnostic}
\label{subsec:validation:pair-response}

The projected statistic $w_p(r_\mathrm{p})$ is built by summing the two-dimensional correlation function on the LOS separation.  A PDF-aware method therefore assigns to each pair not a single inferred LOS separation, but a distribution of possible separations induced by the two radial PDFs.  The purpose of the LOS pair-response diagnostic is to visualize this redistribution directly.  For galaxy pairs whose true projected separation lies inside a narrow interval $[r_{\mathrm{p},\min},r_{\mathrm{p},\max})$, the diagnostic asks how their probability is distributed over the estimated LOS separation $\hat\pi$.

This test deliberately suppresses most of the complexity of a full clustering measurement.  It does not estimate $\xi(r_\mathrm{p},\pi)$ or $w_p(r_\mathrm{p})$, and it does not include random catalogs or estimator normalization.  Instead, it isolates the LOS kernel produced by each PDF representation.  The estimated LOS bins can be chosen as a diagnostic grid or matched to the same $\pi$ grid used in the clustering calculation, so the test applies both to a single wide LOS window and to resolved multi-$\pi$ measurements.

\begin{figure}
    \centering
    \includegraphics[width=1\linewidth]{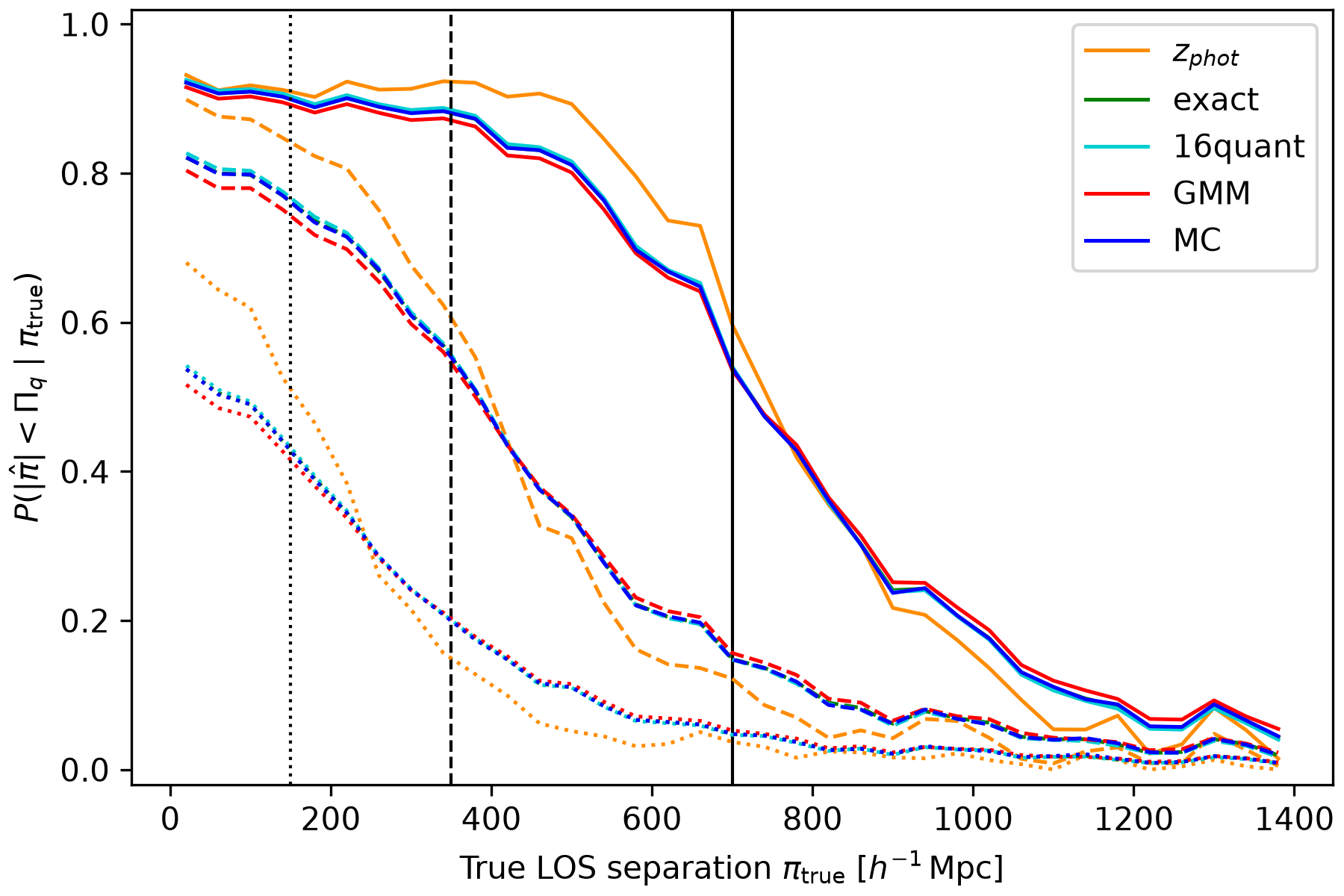}
    \caption{Cumulative LOS inclusion probability derived from the response matrices in Fig.~\ref{fig:kernel_pair_response}. Dotted, dashed, and solid curves correspond to increasing choices of $\Pi_q$ (marked by vertical lines). Curves are shown in different colors for point-estimate ($z_{phot}$), exact, 16quant, gaussian mixture, and Monte Carlo PDF modes.The PDF-aware methods give almost indistinguishable cumulative inclusion probabilities, showing the expected trade-off between close-pair retention and foreground--background leakage as $\Pi_q$ increases.
    }
    \label{fig:los_inclusion}
\end{figure}

\subsubsection{Sampling pairs at fixed true projected separation}

Let $(\alpha_i,\delta_i,z_{i,\rm spec})$ denote the angular coordinates and true redshift of galaxy $i$, and let $\chi_i\equiv \chi(z_{i,\rm spec})$ be its comoving radial distance.  We randomly sample unordered pairs $(i,j)$ from a galaxy sample and retain only those satisfying the chosen cut in true projected separation.  In the implementation used here, $r_{\mathrm{p},\rm true}(i,j)\simeq \bar\chi_{ij}\theta_{ij}$, with $\bar\chi_{ij}=(\chi_i+\chi_j)/2$, and $\pi_{\rm true}(i,j)=|\chi_i-\chi_j|$.

The true LOS axis is partitioned into bins $I_u=[\pi_u,\pi_{u+1})$ and the estimated LOS axis into bins $J_v=[\hat\pi_v,\hat\pi_{v+1})$.  For each selected pair $(i,j)$, let $K_{ij,v}$ be the mass of the method-specific LOS kernel inside the estimated bin $J_v$.  The response matrix is the row average
\begin{equation}
 H_{uv}
 =
 \frac{1}{N_u}
 \sum_{(i,j)\in \mathcal{S}}
 \mathbb{I}\!\left[\pi_{\rm true}(i,j)\in I_u\right]
 K_{ij,v},
\label{eq:pairdiag-Huv}
\end{equation}
where $\mathcal{S}$ is the set of selected pairs and $N_u$ is the number of pairs retained whose true LOS separation falls in row $u$. Each row populated with $H_{uv}$ is therefore the mean conditional kernel in the estimated LOS separation at a fixed true LOS separation.

\subsubsection{Method-specific LOS kernels}

The definition of $K_{ij,v}$ depends on the PDF representation.  For a point-redshift method, the kernel is degenerate and places all probability in the bin containing $|\hat\chi_i-\hat\chi_j|$.  For an empirical grid PDF on a shared radial support $\{\chi_m\}$, the kernel is the distribution of the absolute difference of two draws from the two PDFs, $K^{\rm ePDF}_{ij,v}=\sum_{m,n}p_{im}p_{jn}\,\mathbb{I}[|\chi_m-\chi_n|\in J_v]$.  When the input PDFs are edge-grid histograms, the diagnostic may refine the grid within each input bin to reduce artificial striping in $\hat\pi$.

For the GMM representation, the kernel mass in LOS bin $J_v=[a_v,b_v)$ can be written as
\begin{equation}
 K^{\rm GMM}_{ij,v}
 =
 \sum_{a,c}
 \alpha_{ia}\alpha_{jc}
 \left[
 F_{|X|}\!\left(b_v;m_{iacj},s_{iacj}\right)
 -
 F_{|X|}\!\left(a_v;m_{iacj},s_{iacj}\right)
 \right],
\label{eq:pairdiag-gmm-kernel}
\end{equation}
where $m_{iacj}=\mu_{ia}-\mu_{jc}$, $s_{iacj}^2=\sigma_{ia}^2+\sigma_{jc}^2$, and $F_{|X|}$ is the folded-normal cumulative distribution defined in Sect.~\ref{subsec:methods:gmm}.  For the Monte Carlo mode, $K^{\rm MC}_{ij,v}$ is the empirical fraction of realizations for which $|\chi_i^{(r)}-\chi_j^{(r)}|\in J_v$.  These definitions provide directly comparable LOS kernels for the point-$z$, empirical-PDF, GMM, and MC treatments.

\subsubsection{Inclusion probability into the projected window}

A useful one-dimensional summary of the response matrix is the probability that a pair with true LOS separation in row $u$ ($\pi_u$) is included within a cumulative LOS window $\Pi_q$:
\begin{equation}
 P_{\rm in}(\pi_u;\Pi_q)
 =
 \sum_{v:\,\hat\pi_v^{\rm c}<\Pi_q} H_{uv},
 \qquad \Pi_q\equiv \pi_q,
\label{eq:pairdiag-pin-cumulative}
\end{equation}
where $\hat\pi_v^{\rm c}$ is the center of the estimated LOS bin.  The single-window diagnostic used in a one-bin projected measurement is recovered by setting $\Pi_q=\pi_{\max}$.  For a resolved multi-$\pi$ calculation, the same response matrix can be summarized at several cumulative depths, mirroring the construction of cumulative $w_p(r_\mathrm{p};\Pi_q)$ curves.

\subsubsection{Interpretation and limitations}

In Figure~\ref{fig:kernel_pair_response}, we show the response matrix $H_{uv}$ for each method for a sample of 30 000 galaxy pairs with $z_{\rm spec}$<1.7, randomly selected from DESI.
From Figure~\ref{fig:kernel_pair_response} it can be seen that the empirical-PDF, GMM, and MC treatments produce very similar smooth LOS response matrices.  This agreement indicates that the GMM shell probabilities and the finite MC sampling reproduce the empirical-PDF LOS redistribution well at fixed true $r_\mathrm{p}$. The point-$z$ response is more concentrated and more structured, as expected when the full radial PDF is collapsed to a single scalar estimate.

Figure~\ref{fig:los_inclusion} shows the same information collapsed into cumulative LOS windows. 
Increasing $\Pi_q$ increases the retention of truly close pairs, but also admits a larger fraction of pairs with substantial true LOS separation.  The empirical-PDF, GMM, and MC curves remain nearly indistinguishable across the plotted windows, whereas the point-$z$ treatment follows a different inclusion pattern.  This confirms that the three PDF-aware methods agree closely in their LOS-window probabilities. The same analysis performed in other projected-separation bins over $0.3 \lesssim r_\mathrm{p} \lesssim 15\,h^{-1}{\rm Mpc}$ shows the same qualitative ordering and agreement among the PDF-aware methods. It also emphasizes that the choice of $\pi_{\max}$ controls a trade-off between close-pair retention and foreground--background leakage.

The LOS pair-response diagnostic is a kernel-level test, not a clustering measurement.  Because the sampled pairs are preselected in a narrow interval of true $r_\mathrm{p}$, the response matrix isolates how each method treats the LOS coordinate once the true projected scale is fixed.  This makes it useful for diagnosing LOS retention, leakage, and broadening, and for checking whether the multi-$\pi$ shell probabilities used by different PDF modes are mutually consistent. However, this diagnostic does not include the estimator-level effects that enter the final projected correlation function.  It does not test the behavior of DD, DR, and RR after Landy--Szalay or related normalization, and it does not provide a complete measure of migration across $r_\mathrm{p}$ bins.  Two methods may therefore show similar $H_{uv}$ or $P_{\rm in}(\pi_u;\Pi_q)$ curves while still yielding different $w_p(r_\mathrm{p})$ if one of them transports probability differently across projected-separation bin edges.  This caveat may be relevant for the GMM mode, whose LOS treatment is analytic but whose assignment of $r_\mathrm{p}$ is evaluated in means of component-pairs.  For this reason, the LOS response test should be complemented by the complete clustering-level degradation tests described below.

\subsection{Pair-probability calibration}
\label{sec:pair_probability_calibration}

The line-of-sight response described above characterizes how photometric-redshift uncertainty redistributes pair counts along the radial direction. A complementary question is whether the full redshift PDFs carry useful pair-level information that is lost when each galaxy is compressed to a single point estimate. We therefore introduce a simple calibration test based on the probability that a projected pair is truly close along the line of sight. The purpose of this test is not to measure a correlation function directly, but to ask whether the PDF-based pair weights can be interpreted as meaningful probabilistic pair-membership information.

Calibration tests for probabilistic photo-$z$ estimates commonly ask whether quoted probabilities or credible regions correspond to their empirical frequencies in a validation sample \citep{wittman2016,schmidt2020}. Here we apply the same principle to a derived pair event: whether the two galaxies lie within a specified true LOS separation.

We use the CosmoDC2 galaxies with $0.1<z_{\rm true}<0.7$ and select projected pairs in the range $1<r_\mathrm{p}<5\,h^{-1}{\rm Mpc}$.  For each pair $(i,j)$ we define the true close-pair indicator
\begin{equation}
T_{ij}(\Pi_{\max}) =
\begin{cases}
1, & |\chi_i^{\rm true}-\chi_j^{\rm true}| < \Pi_{\max},\\
0, & |\chi_i^{\rm true}-\chi_j^{\rm true}| \geq \Pi_{\max},
\end{cases}
\label{eq:true_pair_indicator}
\end{equation}
where $\chi^{\rm true}$ is the comoving distance corresponding to the true redshift.  For the full-PDF case we compute the probability that the same pair satisfies this condition by marginalizing over the two redshift PDFs,
\begin{equation}
P_{ij}^{\rm PDF}(\Pi_{\max}) =
\int d\chi_i\,d\chi_j\,
 p_i(\chi_i)\,p_j(\chi_j)\,
 \Theta\left(\Pi_{\max}-|\chi_i-\chi_j|\right),
\label{eq:pdf_pair_probability}
\end{equation}
where $\Theta$ is the Heaviside step function.  In contrast, the point-photo-$z$ estimate produces only a hard decision,
\begin{equation}
P_{ij}^{\rm point}(\Pi_{\max}) =
\begin{cases}
1, & |\chi_i^{\rm phot}-\chi_j^{\rm phot}| < \Pi_{\max},\\
0, & |\chi_i^{\rm phot}-\chi_j^{\rm phot}| \geq \Pi_{\max}.
\end{cases}
\label{eq:point_pair_decision}
\end{equation}
We then bin the pairs by $P_{ij}^{\rm PDF}$ and measure the mean value of $T_{ij}$ in each bin.  A perfectly calibrated probabilistic pair assignment would follow $\langle T_{ij}\rangle = P_{ij}^{\rm PDF}$.

\begin{figure}
    \centering
    \includegraphics[width=1\linewidth]{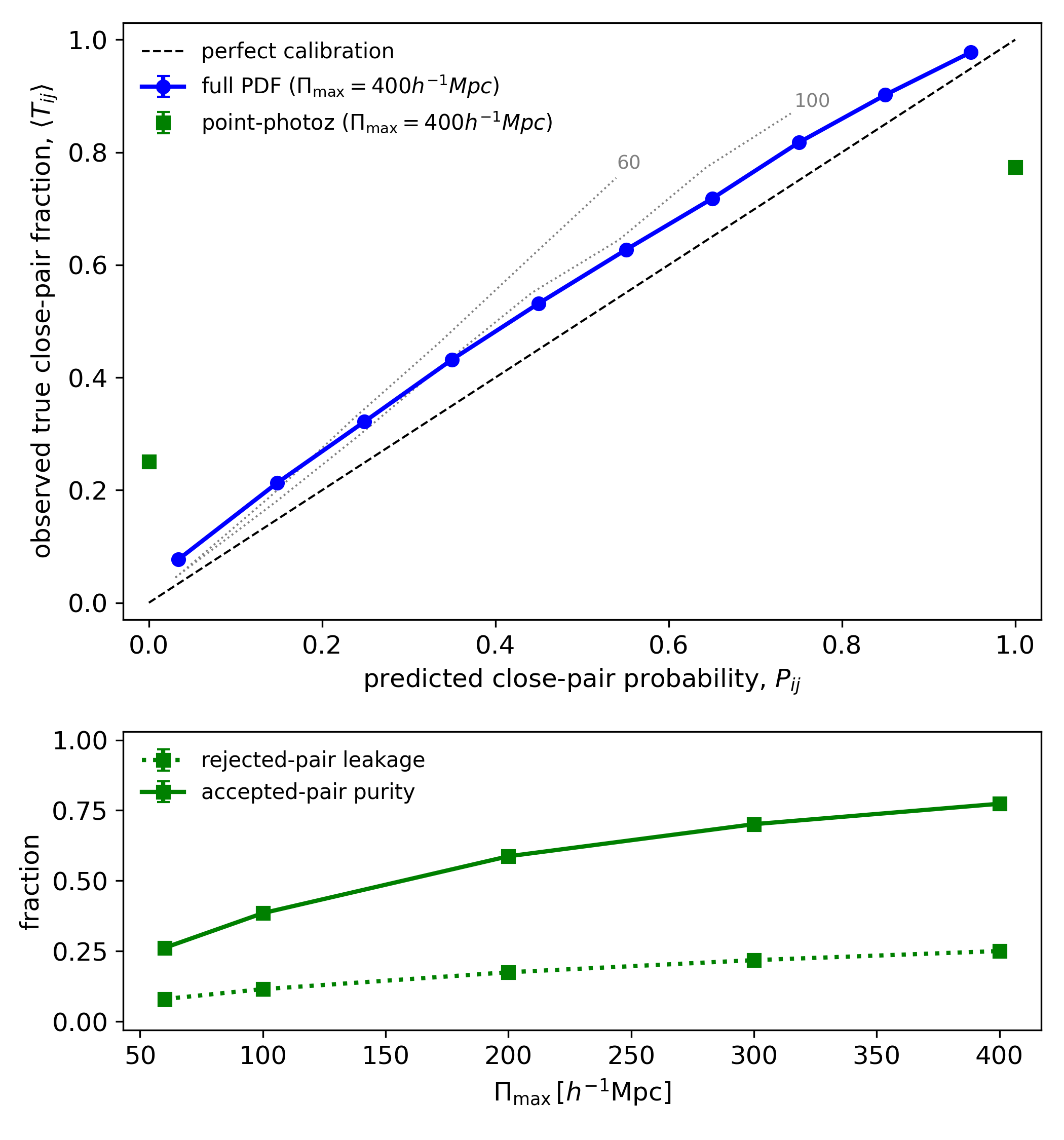}
    \caption{Pair-level calibration test for CosmoDC2 galaxies with $0.1<z_{\rm true}<0.7$ and $1<r_\mathrm{p}<5\,h^{-1}{\rm Mpc}$.  The upper panel compares the PDF-predicted close-pair probability with the observed fraction of pairs that are truly within $\Pi_{\max}$ along the line of sight.  The blue curve shows the full-PDF result for $\Pi_{\max}=400\,h^{-1}{\rm Mpc}$, while gray dotted curves show smaller integration limits for reference.  The green points show the binary point-photo-$z$ decisions at the same $\Pi_{\max}=400\,h^{-1}{\rm Mpc}$.  The lower panel summarizes the point-photo-$z$ classification as a function of $\Pi_{\max}$, showing the rejected-pair leakage, $P(T_{ij}=1\,|\,P_{ij}^{\rm point}=0)$, and the accepted-pair purity, $P(T_{ij}=1\,|\,P_{ij}^{\rm point}=1)$.}
    \label{fig:pair_probability_calibration}
\end{figure}

The results of this test are shown in Figure~\ref{fig:pair_probability_calibration}. The upper panel compares the observed true close-pair fraction, $\langle T_{ij}\rangle$, with the predicted close-pair probability.  The blue curve shows the full-PDF result for $\Pi_{\max}=400\,h^{-1}{\rm Mpc}$, close to the recovery scale identified from the cumulative projected-correlation signal for this sample (see Fig. \ref{fig:deg_cumwp_full}). The gray dotted curves show the same full-PDF diagnostic for smaller integration limits, $\Pi_{\max}=60$ and $100\,h^{-1}{\rm Mpc}$, as references.  The green points show the corresponding point-photo-$z$ hard decisions for $\Pi_{\max}=400\,h^{-1}{\rm Mpc}$. Of course, point estimates can only classify pairs as rejected, $P_{ij}^{\rm point}=0$, or accepted, $P_{ij}^{\rm point}=1$.

This upper panel should be read as a probability-calibration diagram. For example, a point on the blue curve at $P_{ij}^{\rm PDF}\simeq 0.45$, means that the full-PDF calculation assigns those pairs a $\simeq45\%$ probability of lying within the adopted line-of-sight aperture. The corresponding ordinate shows that, in the mock catalog, approximately
$53\%$ of those same pairs are actually close according to their true comoving distances.  The full-PDF probabilities are therefore not arbitrary weights: they provide an ordered and nearly calibrated ranking of pair membership. The fact that the blue curve lies slightly above the one-to-one relation indicates that the PDF overlap probability is somewhat conservative for these small-$r_\mathrm{p}$ pairs. This is expected, at least in part, because $P_{ij}^{\rm PDF}$ is computed from the product of the two individual redshift PDFs (see Eq.~\eqref{eq:pdf_pair_probability}) and does not include the additional prior information that galaxies close on the sky are themselves clustered. The important result is that the full-PDF curve is monotonic and lies close to the one-to-one relation, showing that the PDF overlap probability provides a meaningful ranking of pair membership: pairs assigned larger $P_{ij}^{\rm PDF}$ are more likely to be truly close in the simulation.

The contrast with the point-photo-$z$ case is more severe. A point estimate collapses the same information into a binary decision: a pair is either accepted as close, $P_{ij}^{\rm point}=1$, or rejected, $P_{ij}^{\rm point}=0$. For $\Pi_{\max}=400\,h^{-1}{\rm Mpc}$, pairs classified as not close by the point-photo-$z$ estimate still have a true-close fraction of $\simeq\,25\%$. Conversely, pairs classified as close by the point-photo-$z$ estimate are truly close only about $77\%$ of the time. In plain language, the point-photo-$z$ selection simultaneously misses a substantial population of genuine close pairs and admits a significant population of interlopers. The lower panel quantifies this behavior as a function of $\Pi_{\max}$. The rejected-pair leakage, $P(T_{ij}=1\,|\,P_{ij}^{\rm point}=0)$, increases from $\simeq\,8\%$ at $\Pi_{\max}=60\,h^{-1}{\rm Mpc}$ to about $25\%$ at
$\Pi_{\max}=400\,h^{-1}{\rm Mpc}$. The accepted-pair purity, $P(T_{ij}=1\,|\,P_{ij}^{\rm point}=1)$, increases from only $\simeq26\%$ to $\simeq77\%$ over the same range. Increasing the integration scale therefore makes the hard point-photo-$z$ decision less catastrophic, but it
does not remove the information loss caused by compressing each redshift PDF into a single radial coordinate.

This test clarifies the sense in which full PDFs add information to the clustering measurement. They do not make the radial positions of individual galaxies intrinsically more precise, nor should they necessarily produce a sharper line-of-sight signal.  Instead, they replace overconfident hard pair assignments with continuous probabilities that track the true close-pair fraction.  The improvement is therefore a gain in probabilistic fidelity: full PDFs preserve partial pair-membership information that is discarded by any scalar photo-$z$ compression.

\subsection{Controlled degradation test}
\label{subsec:validation:degradation}

The LOS pair-response diagnostic in Section~\ref{subsec:validation:pair-response} shows how each method redistributes pair probability along the line of sight at fixed true $r_\mathrm{p}$, but it does not measure the impact of that redistribution on the clustering estimator itself. To quantify the effect on the final projected signal, we perform a controlled degradation experiment using the CosmoDC2 sample, where the true redshifts are known and the clustering signal can be measured before imposing photometric-redshift uncertainty.

\begin{figure}
    \centering
    \includegraphics[width=1\linewidth]{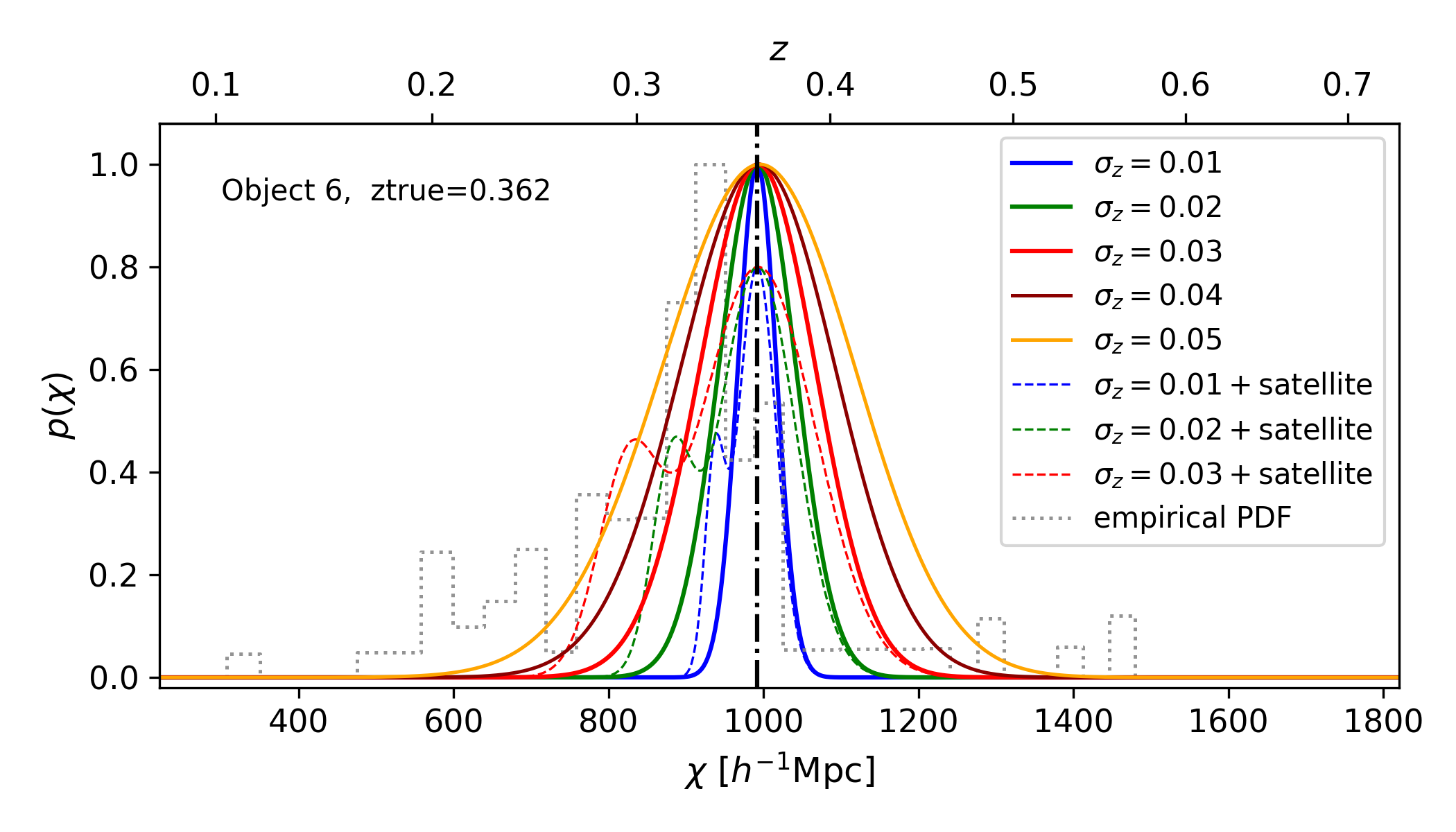}
    \caption{Gaussian surrogate PDFs for a test CosmoDC2 galaxy at $z_{\rm true}=0.362$. Solid lines show single Gaussians of increasing width, dashed lines show two-component Gaussian surrogates, and the dotted histogram shows the corresponding empirical PDF.}
    \label{fig:gauss}
\end{figure}

For $\sim 674\,000$ galaxies at $0.1<z_{\rm true}<0.7$, the experiment replaces each true redshift by a sequence of increasingly realistic photometric-redshift PDFs. We first use single Gaussian PDFs centered on $z_{\rm true}$ with fixed widths $\sigma_z=0.01$, $0.02$, $0.03$, $0.04$, and $0.05$. This provides a clean test of radial smearing alone, with no central redshift bias and no multimodality. We then repeat the analysis with two-component Gaussian surrogates, consisting of a dominant peak at $z_{\rm true}$ plus a weaker, shifted component, to introduce controlled non-Gaussian structure. Figure~\ref{fig:gauss} illustrates these surrogate PDFs for one galaxy, together with its empirical PDF.

For each PDF family, we compare the truth-redshift result with the MC, 16quant, and GMM PDF-aware modes, and with the exact empirical-PDF mode where computationally feasible. Comparisons are performed in resolved multi-$\pi$ bins, so that we can inspect both $\xi(r_\mathrm{p},\pi)$ and the cumulative projected signal $w_p(r_\mathrm{p};\Pi)$ as a function of LOS depth. This makes it possible to identify where the clustering signal is redistributed in $\pi$, how much signal is recovered by a given $\Pi$, and whether differences between methods arise mainly from LOS smearing, projected-bin migration, or the internal PDF approximation.

This scheme separates three effects. The fixed-width Gaussian case isolates the response to a known radial uncertainty scale. The two-component surrogate tests sensitivity to asymmetric or multi-component PDFs while retaining full control over the input model. The empirical-PDF case then shows the combined effect of realistic PDF widths, irregular shapes, tails, and secondary peaks.

\begin{figure}
    \centering
    \includegraphics[width=1\linewidth]{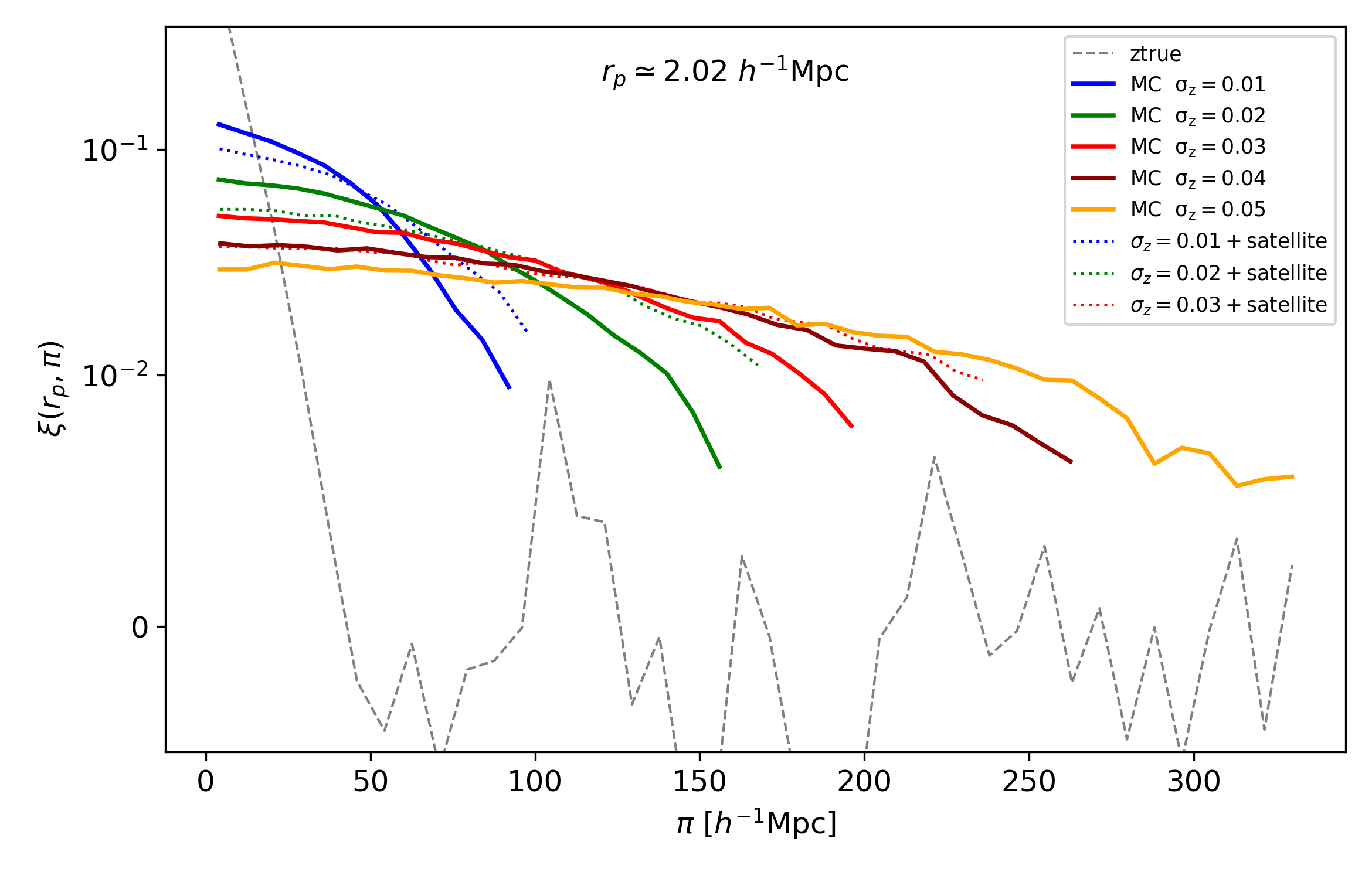}
    \caption{Clustering signal $\xi(r_\mathrm{p},\pi)$ for the MC method as a function of $\pi$, evaluated at $r_\mathrm{p}\simeq 2\,h^{-1}{\rm Mpc}$. The truth-redshift curve declines rapidly at small LOS separation, while the Gaussian and two-component surrogate PDFs progressively smear the signal to larger $\pi$.}
    \label{fig:xi_vs_pi}
\end{figure}

Figure~\ref{fig:xi_vs_pi} shows the effect of this degradation directly on $\xi(r_\mathrm{p},\pi)$ for the MC mode at $r_\mathrm{p}\simeq 2\,h^{-1}{\rm Mpc}$. The truth-redshift signal is sharply concentrated at small LOS separation and falls rapidly beyond $\pi\sim 30$--$40\,h^{-1}{\rm Mpc}$. As $\sigma_z$ increases, the same clustering signal is redistributed over a much broader range of $\pi$: the peak amplitude decreases, while a long, low-amplitude tail appears at large LOS separation. The two-component surrogates produce the same qualitative behavior, with additional broadening associated with the satellite component. Thus, the primary effect of the PDF treatment is not to create a new clustering signal, but to move the existing close-pair signal outward in LOS separation.

\begin{figure*}
    \centering
    \includegraphics[width=0.9\linewidth]{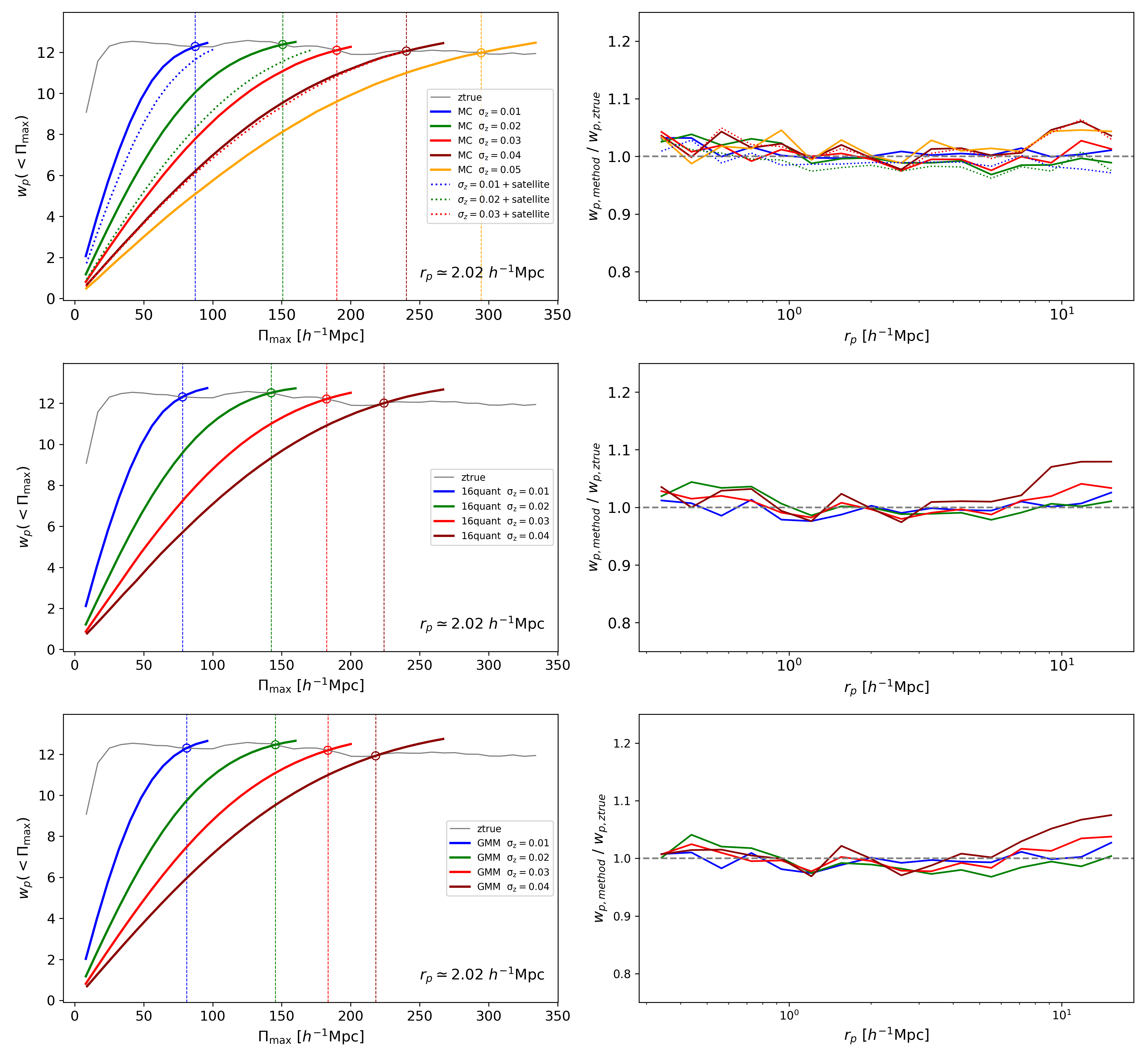}
    \caption{Cumulative projected signal analysis with the full empirical PDFs for CosmoDC2 galaxies at $0.1<z_{\rm true}<0.7$. Rows show, from top to bottom, the MC, 16quant, and GMM modes. The left and right panels show the corresponding $w_p$ and $w_p/w_{p,z_{true}}$ estimates.} 
    \label{fig:deg_cumwp}
\end{figure*}

The cumulative view in Figure~\ref{fig:deg_cumwp} makes this redistribution clearer. 
In Figure~\ref{fig:deg_cumwp}, we show the cumulative projected signal $w_p (<\Pi_{\max})$ as a function of LOS integration depth $\Pi_{\max}$, for the truth-redshift measurement and for each degraded PDF case, at fixed $r_\mathrm{p}\simeq 2\,h^{-1}{\rm Mpc}$, with rows corresponding to the MC, 16quant, and GMM modes.
The truth-redshift curve reaches its projected asymptotic signal at relatively small $\Pi$, since all of the real clustering signal is captured within a narrow LOS window. The PDF-aware curves rise more slowly because part of the probability that belongs to physically close pairs has been displaced to larger LOS separations. For each degraded sample, we define an operational recovery depth, $\Pi_{\max}^{\rm rec}$, as the first intersection between the cumulative PDF-aware curve and the truth-redshift cumulative curve. Since the truth-redshift curve is nearly flat at these depths, this is effectively the LOS scale required to recover the projected truth signal at that $r_\mathrm{p}$.

The right panels of Fig.~\ref{fig:deg_cumwp} show that, when evaluated at this recovery scale, the recovered $w_p(r_\mathrm{p})$ is close to the truth-redshift result over the full projected-separation range. The ratios are nearly flat and generally remain within a few percent of unity for the single-Gaussian tests. The MC, 16quant, and GMM modes show the same overall behavior, although the GMM sweep is restricted to the narrower Gaussian cases shown in the figure. The two-component surrogates also recover the truth signal well once the larger effective LOS width is included. These results indicate that the PDF-aware methods are behaving consistently: the degradation is mainly a controlled LOS redistribution of the signal, and the projected clustering amplitude can be recovered provided that the integration depth is matched to the pairwise radial smearing.

\begin{figure}
    \centering
    \includegraphics[width=1\linewidth]{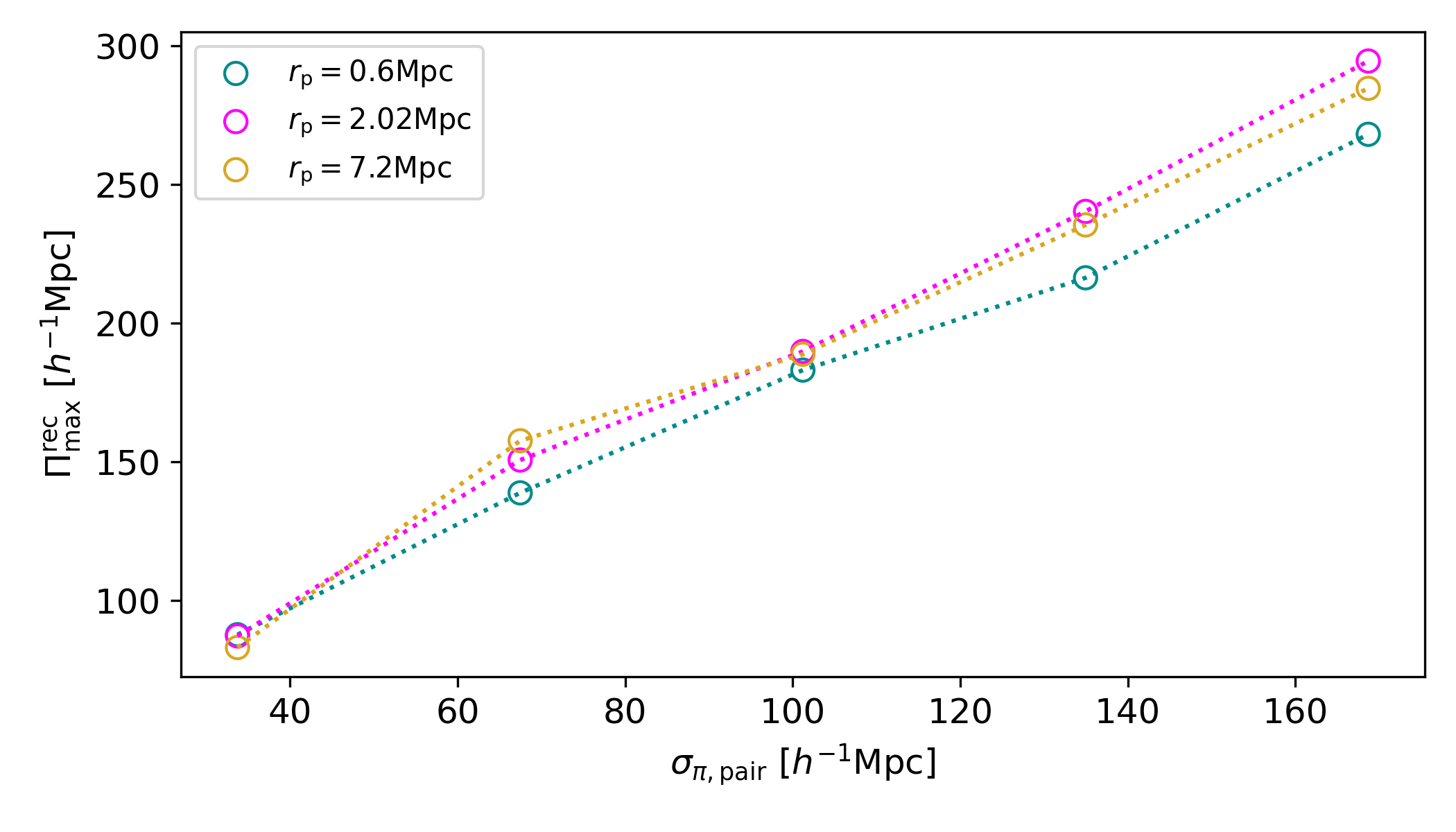}
    \caption{Recovery depth $\Pi_{\max}^{\rm rec}$ as a function of the effective pairwise LOS dispersion $\sigma_{\pi,{\rm pair}}$ for three projected-separation bins. The nearly linear trend shows that the integration depth needed to recover the truth-redshift projected signal is set primarily by the pairwise radial smearing.}
    \label{fig:pirec_vs_sigma}
\end{figure}

Figure~\ref{fig:pirec_vs_sigma} summarizes this calibration by plotting $\Pi_{\max}^{\rm rec}$ against the effective pairwise LOS dispersion, $\sigma_{\pi,{\rm pair}}$, for three projected-separation bins.
The relation is approximately linear and only weakly dependent on $r_\mathrm{p}$ over the range shown. This is a useful empirical result: the appropriate LOS integration scale is not arbitrary but tracks the pairwise radial uncertainty of the sample. This also explains why broader PDFs require much larger projected windows, even when they are unbiased in their central redshift. A fixed value of $\Pi_{\max}$ can therefore be too small for broad PDFs, while an excessively large value will admit unnecessary foreground--background leakage and increase the noise.

\begin{figure*}
    \centering
    \includegraphics[width=1\linewidth]{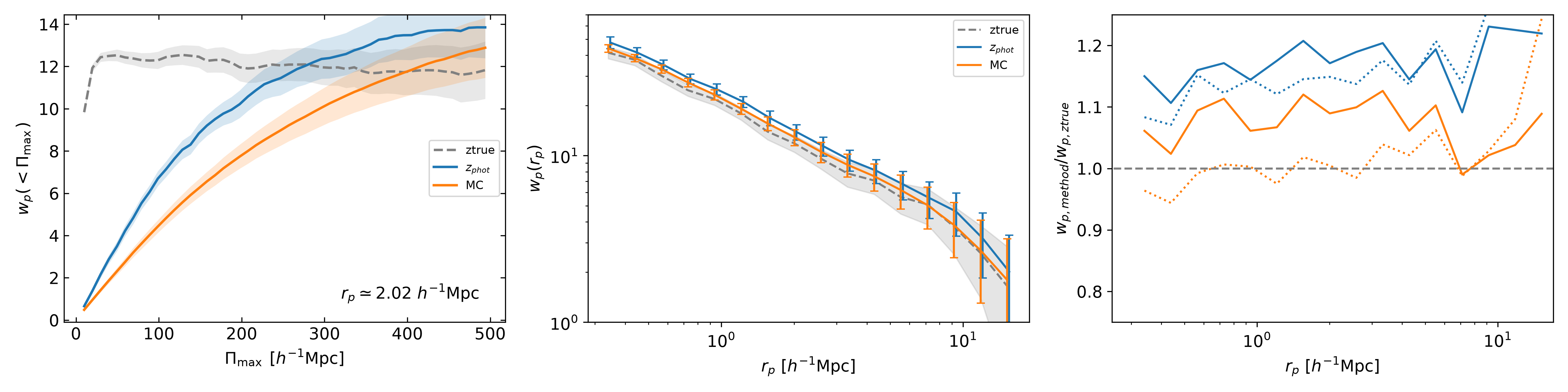}
    \caption{Cumulative projected signal analysis with full PDFs for CosmoDC2 galaxies with $0.1<z_{\rm true}<0.7$. The middle and right panels show the corresponding $w_p$ and $w_p/w_{p,z_{true}}$ estimates (solid lines), when integrating up to $\Pi_{\rm max}=2\Pi_{\rm max}^{\rm guess}$. The dashes lines show the ratio when $\Pi_{\rm max}=\Pi_{\max}^{\rm rec}$.}
    \label{fig:deg_cumwp_full}
\end{figure*}

Empirical PDFs provide a more realistic and less idealized case. Figure~\ref{fig:deg_cumwp_full} compares the truth-redshift, point-estimate $z_{\rm phot}$, and full-PDF MC measurements. The cumulative curve for the full-PDFs MC case rises more slowly than the point-estimate curve, reflecting the broad and irregular empirical PDFs, but approaches the projected signal with the truth-redshift when integrated to sufficiently large $\Pi_{\max}$. At the conservative choice $\Pi_{\max}=2\Pi_{\max}^{\rm guess}$, the MC result is closer to the truth measurement than the point-estimate result, although the ratio still shows residual scale-dependent excess caused by the very large LOS window. When, instead, it is evaluated at the recovery depth inferred from the cumulative curve, the MC ratio becomes substantially flatter and closer to unity. The point-estimate measurement, by contrast, remains more biased because collapsing each PDF to a single redshift does not represent the full LOS probability distribution.

This comparison also exposes a practical limitation. In a controlled mock such as CosmoDC2, the gray $z_{\rm true}$ curve is known, and $\Pi_{\max}^{\rm rec}$ can be identified by the intersection with the degraded cumulative curve. In a real large photometric survey, such as LSST, that reference curve is precisely the unknown quantity. The recovery depth therefore cannot be read directly from the data. Instead, it should be calibrated from the PDFs themselves, from realistic simulations, from spectroscopic validation subsamples, or from stability tests of $w_p(r_\mathrm{p};\Pi)$ as the LOS window is enlarged. The cumulative curves remain useful, but their role changes: without $z_{\rm true}$ they diagnose convergence and leakage, rather than providing an exact recovery point.

The controlled degradation experiment leads to three conclusions. First, PDF-aware projected clustering is fundamentally a redistribution problem: radial uncertainty spreads close-pair signal along the LOS, and the relevant question is how far one must integrate to recover that redistributed signal. Second, the MC, 16quant, and GMM implementations give mutually consistent recovery behavior in the regimes tested, supporting the interpretation that they approximate the same PDF-marginalized pair-count target. Third, the integration depth should be tied to the effective pairwise smearing of the sample and checked with cumulative $w_p(r_\mathrm{p};\Pi)$ curves. In controlled Gaussian tests this produces a well-defined $\Pi_{\max}^{\rm rec}$ and nearly unbiased recovery of $w_p(r_\mathrm{p})$; in the full empirical-PDF case it remains a practical diagnostic for balancing signal recovery against foreground--background leakage.

\subsection{Parameter convergence of the PDF approximations}
\label{sec:param_convergence}

The previous section tested how radial uncertainty redistributes the projected clustering signal and how the choice of LOS integration depth affects its recovery. We now analyze how sensitive are the PDF-aware measurements to the internal resolution parameters of each approximation. This test is performed with the full empirical PDFs of the same CosmoDC2 sample at $0.1<z_{\rm true}<0.7$, but using a random subsample of $\sim 210\,000$ data galaxies and $\sim 840\,000$ random galaxies in order to keep the convergence runs computationally tractable. For each method, we compare $w_p(r_\mathrm{p})$ with the corresponding truth-redshift measurement and summarize the differences with the fractional RMS and the median fractional bias over the plotted $r_\mathrm{p}$ range.

\begin{figure}
    \centering
    \includegraphics[width=1\linewidth]{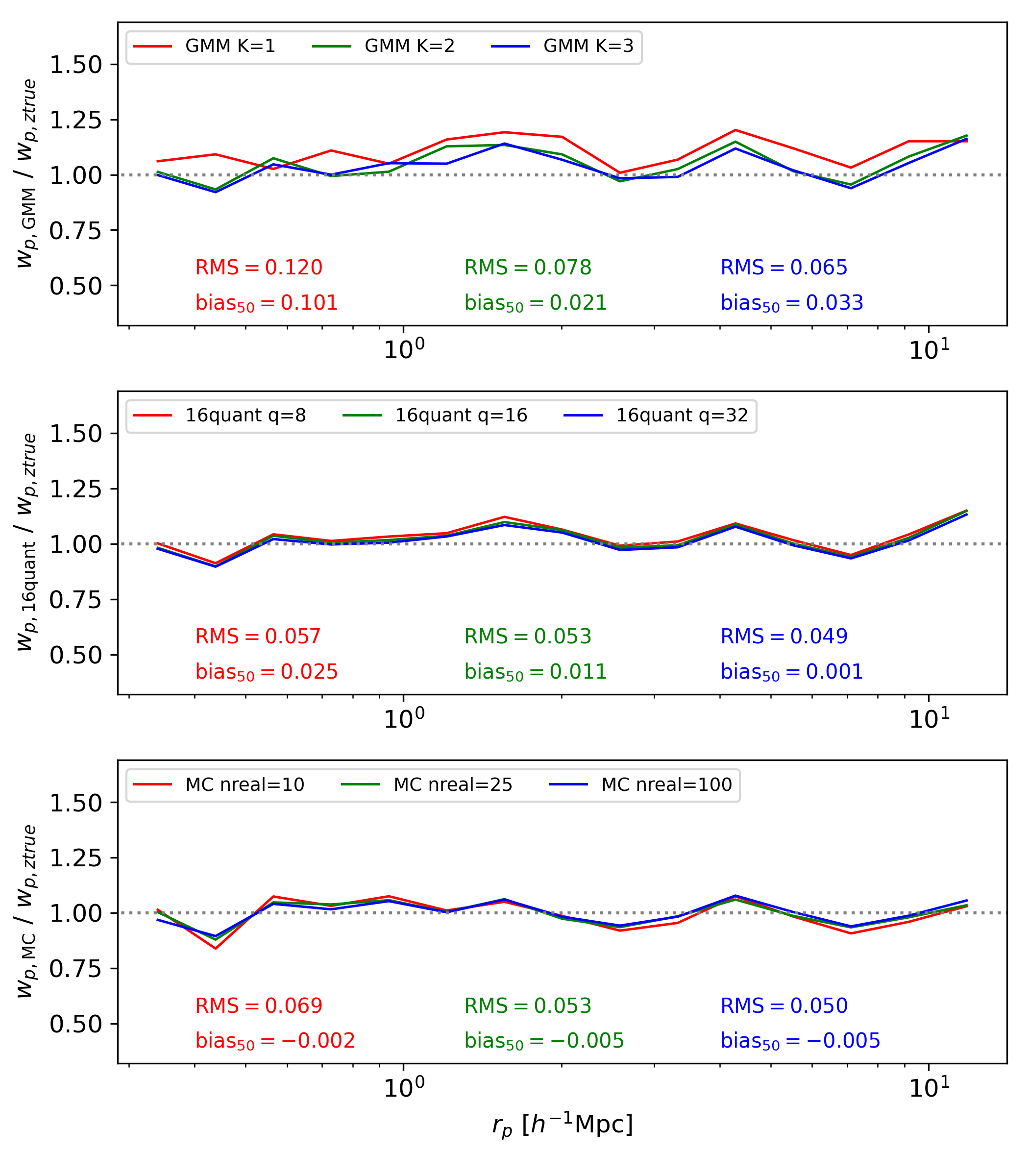}
    \caption{Convergence of the main PDF-aware approximations with their internal resolution parameters for a subsample of CosmoDC2 galaxies with full empirical PDFs. The panels show $w_p/w_{p,z_{\rm true}}$ for GMM with different numbers of components $K$, 16quant with different numbers of equal-probability nodes $q$, and MC with different numbers of realizations $n_{\rm real}$. The annotations give the fractional RMS relative to the truth-redshift measurement and the median fractional bias over the plotted $r_\mathrm{p}$ range.}
    \label{fig:param_wpratios}
\end{figure}

Figure~\ref{fig:param_wpratios} shows the response of the three approximate PDF modes to their main control parameters. For the GMM mode, increasing the number of components from $K=1$ to $K=2$ produces the largest improvement: the fractional RMS decreases from $0.120$ to $0.078$, and the median bias decreases from $0.101$ to $0.021$. Increasing to $K=3$ gives a further, but smaller, reduction in RMS to $0.065$. This behavior is expected for irregular empirical PDFs: a single Gaussian captures only the gross width of the radial uncertainty, whereas two or three components can represent asymmetry and secondary structure more faithfully. The remaining offsets are consistent with the fact that the current GMM implementation treats the LOS probability analytically but assigns $r_\mathrm{p}$ from component-pair means.

The 16quant mode shows weaker parameter dependence. Even $q=8$ equal-probability nodes give a fractional RMS of $0.057$ and a median bias of $0.025$. Increasing the resolution to $q=16$ reduces these values to $0.053$ and $0.011$, while $q=32$ gives a modest additional improvement to $0.049$ and an almost negligible median bias. Thus, most of the information needed for the projected statistic is already captured by a relatively small quantile representation. The improvement with $q$ is monotonic but saturates rapidly, making $q=16$ a useful compromise between accuracy and cost. Since $q$ sets the quadrature resolution of each individual PDF, rather than that of the galaxy sample, it is not directly determined by the survey volume or number of pairs. A larger catalog may nevertheless require tighter convergence because of its smaller statistical uncertainties, while different redshift intervals may require reassessment if the PDF shapes change significantly. The optimal number depends on both the PDF ensemble and the downstream science metric \citep{malz2018}. We therefore regard the adopted $q$ value as an empirical default rather than a universal choice.

The MC curves show the expected convergence with the number of realizations. With $n_{\rm real}=10$ the RMS is $0.069$, while $n_{\rm real}=25$ reduces it to $0.053$. Increasing to $n_{\rm real}=100$ produces only a small additional gain, with RMS $0.050$. The median bias is already close to zero for all three choices, indicating that the dominant effect of increasing $n_{\rm real}$ is to reduce realization noise. In this configuration, $n_{\rm real}\simeq25$ is therefore sufficient for percent-level convergence of the mean projected signal, while larger values mainly provide smoother curves.

Overall, Fig.~\ref{fig:param_wpratios} shows that the three approximations converge toward a common PDF-aware result as their internal resolution is increased. The convergence is fastest for 16quant and MC, which both reach low fractional RMS values for practical parameter choices. The GMM approximation also improves substantially with $K$, but requires at least two components for the full empirical PDFs considered here and remains somewhat more sensitive to the details of PDF compression. These results motivate the default choices used in the following applications: $q\simeq16$ for 16quant, $n_{\rm real}\simeq25$ for MC, and $K=2$--$3$ for GMM when the additional cost is acceptable. An independent reduced-sample benchmark against the exact empirical-PDF calculation is presented in Appendix~\ref{app:exact_benchmark}. This test confirms that the practical PDF approximations reproduce the deterministic
PDF-marginalized projected measurement within the accuracy reported there, while requiring substantially lower computational cost.

\subsection{Application to real catalogs: DESI red and blue galaxy clustering}
\label{sec:desi_red_blue}

We finally apply the PDF-aware framework to an observed galaxy sample, using DESI Bright Galaxy Survey (BGS) objects with $0.05<z_{\rm spec}<0.4$ selected from the footprint described in Sect.~\ref{sec:DESI}. The galaxies are divided into red and blue populations with an empirical diagonal boundary in the $(r-z)$--$(z-W1)$ color plane, chosen to approximately separate the two modes of the $r-z$ distribution. This division is sufficient for the methodological purpose of the experiment, but it should not be interpreted as a controlled measurement of the intrinsic color dependence of clustering: the red sample extends to somewhat larger distances and the two populations therefore have different redshift distributions, luminosity mixtures, and radial-PDF properties. Instead, we ask whether the full-PDF estimator can reproduce the relative clustering measured from the spectroscopic redshifts for two realistic populations with markedly different clustering amplitudes.

\begin{figure*}
    \centering
    \includegraphics[width=1\linewidth]{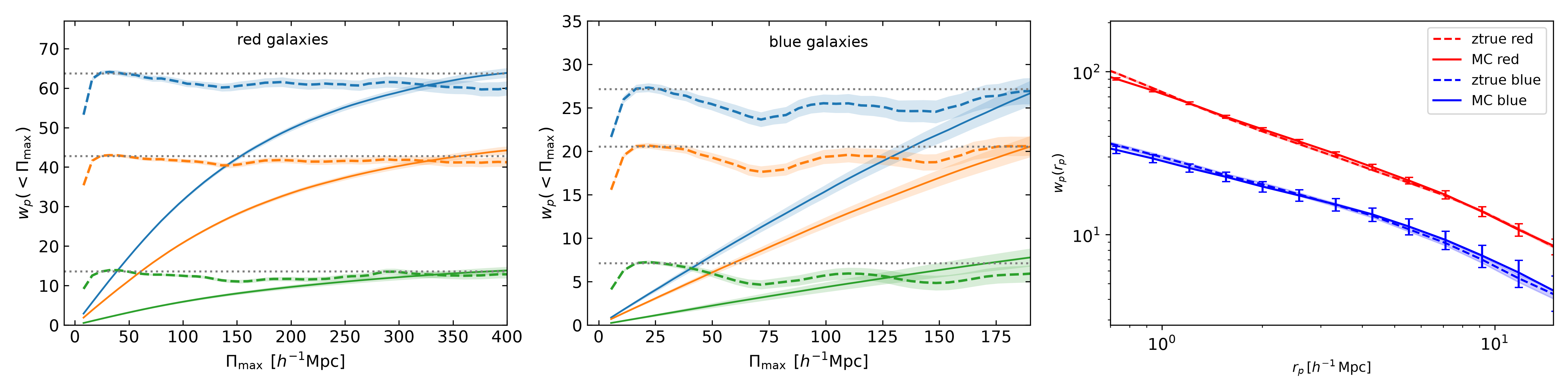}
    \caption{Cumulative projected signal for DESI BGS galaxies with $0.05<z_{\rm spec}<0.4$. The left and middle panels show $w_p(<\Pi_{\max})$ for red and blue galaxies, respectively, using spectroscopic redshifts (dashed) and Monte Carlo sampling of the full PDFs (solid) at three representative $r_\mathrm{p}$ bins ($\sim$1.2, $\sim$2 and $\sim$9.2$\,h^{-1}\mathrm{Mpc}$ from top to bottom, respectively). Dotted lines mark the spectroscopic plateau levels. The right panel compares $w_p(r_\mathrm{p})$ obtained with $\Pi_{\max}=40\,h^{-1}\mathrm{Mpc}$ for the spectroscopic measurements and the recovery limits $\Pi_{\max}=400$ and $180\,h^{-1}\mathrm{Mpc}$ for the red and blue MC measurements. The agreement is good and reproduces both the shape and the relative amplitudes of the red and blue spectroscopic clustering.}
    \label{fig:desi_red_blue}
\end{figure*}

Figure~\ref{fig:desi_red_blue} shows the cumulative projected signal for the red and blue samples. The spectroscopic curves reach their principal plateau by $\Pi_{\max}\simeq40\,h^{-1}\mathrm{Mpc}$, after which the cumulative integral is increasingly affected by low-amplitude fluctuations in $\xi(r_\mathrm{p},\pi)$. We therefore use this early plateau as the spectroscopic reference rather than integrating the precise redshifts over several hundred megaparsecs. By contrast, the MC curves rise gradually because the empirical PDFs redistribute physically correlated pairs over a much broader range of apparent LOS separations. The corresponding recovery depths are sample dependent: the red PDFs require $\Pi_{\max}\simeq400\,h^{-1}\mathrm{Mpc}$, whereas the blue sample reaches the spectroscopic plateau at approximately $180\,h^{-1}\mathrm{Mpc}$. The similar intersection scales obtained in the three representative $r_\mathrm{p}$ bins indicate that a single recovery depth provides a useful description within each population, while the large difference between the red and blue values demonstrates that $\Pi_{\max}$ cannot be considered universal. These realistic full-PDF measurements remain practical with modest hardware. In a 15-core notebook, the MC calculations required $\sim$9 and $\sim$15 min for the blue and red samples, respectively, using well resolved LOS bins of width $\Delta\pi \simeq 6 - 8\,h^{-1}\,\mathrm{Mpc}$ and including estimation of bootstrap uncertainties. Controlled timing comparisons among the different PDF-aware modes are presented in Appendix B.

The smallest projected separations require additional caution. Count-level diagnostics show a pronounced deficit of angularly close data--data pairs in the spectroscopic sample, producing a steep decline of the red cumulative signal below $r_\mathrm{p}\simeq0.8\,h^{-1}\mathrm{Mpc}$. This behavior is consistent with DESI fiber-assignment incompleteness, which is intrinsically pair dependent and becomes most severe in regions of high projected target density. BGS simulations show that the assignment completeness can be very low in the centers of massive clusters \citep{smith2019}. The effect is consequently expected to be stronger for the more highly clustered red population. More recent DESI analyses also find a scale-dependent suppression of two-point clustering and require pairwise or angular mitigation schemes to recover the complete-sample statistic \citep{bianchi2025}. Since no inverse pair probability correction is applied here, we exclude $r_\mathrm{p}\lesssim0.8\,h^{-1}\mathrm{Mpc}$ from the recovery assessment. The PDF methods propagate radial uncertainty, but they cannot restore galaxy pairs absent from the input spectroscopic catalog.

The blue spectroscopic curves show a moderate decrease followed by a later rise as $\Pi_{\max}$ is enlarged. Because $\partial w_p(<\Pi)/\partial\Pi=2\xi(r_\mathrm{p},\Pi)$, this behavior corresponds to alternating low-amplitude negative and positive LOS contributions after the main signal has accumulated. Fiber assignment may contribute to the downturn at small $r_\mathrm{p}$, but it is unlikely to explain the common structure at all three transverse separations, particularly at $r_\mathrm{p}\simeq9\,h^{-1}\mathrm{Mpc}$. We therefore interpret the non-monotonicity more conservatively as a combination of correlated sample fluctuations, redshift-space anisotropy, residual spectroscopic selection effects, and the lower signal-to-noise of the blue population. The MC calculation smooths this fine LOS structure through the radial PDFs. Its relevant success criterion is recovery of the stable projected amplitude, rather than reproduction of every fluctuation in the spectroscopic cumulative curve.

The right panel of Fig.~\ref{fig:desi_red_blue} summarizes the final comparison. Using $\Pi_{\max}=40\,h^{-1}\mathrm{Mpc}$ for the spectroscopic reference and the sample-specific recovery depths for the MC measurements, the full-PDF results reproduce both the shapes of the red and blue $w_p(r_\mathrm{p})$ curves and their large relative amplitude for $r_\mathrm{p}\gtrsim0.8\,h^{-1}\mathrm{Mpc}$. This agreement is non-trivial because the two samples differ simultaneously in redshift distribution, clustering strength, and PDF structure. These results therefore support the main conclusion of the controlled tests in a realistic setting: pair-level PDF propagation can recover the projected clustering contrast of observed galaxy populations, provided that the LOS integration scale is calibrated for each sample and that angular completeness systematics are treated independently through appropriate weights or conservative scale cuts.

\section{Conclusions}\label{sec:conclusions}
We have presented \scalebox{1.2}{$\nu$}\textsc{Gundam}\xspace, a unified, high-performance pair-counting framework for point-redshift and PDF-aware spatial correlation functions. The code extends the original \textsc{Gundam} architecture to projected measurements, in which the distance to each galaxy is represented either by a scalar value or by a radial probability distribution. In the PDF-aware case, the primary output is not only a projected statistic, but also a set of expected DD, DR, and RR count fields on the two-dimensional $(r_\mathrm{p},\pi)$ grid. The projected correlation function $w_p(r_\mathrm{p})$ is then obtained by summing the corresponding estimator over a chosen LOS integration depth. The distinctive contribution of \scalebox{1.2}{$\nu$}\textsc{Gundam}\xspace is to expose several interchangeable PDF-aware projected pair-counting algorithms within a single public estimator and resampling interface, with compiled kernels designed for catalogs containing millions of PDFs.

The framework implements four complementary treatments of radial uncertainty. The exact empirical-PDF mode evaluates the PDF-marginalized pair weights on a common radial grid and provides the deterministic reference calculation where feasible. The GMM mode compresses each empirical PDF into a small number of Gaussian components and evaluates LOS shell probabilities analytically. The MC mode averages standard point-redshift pair counts over repeated draws from the input PDFs. Finally, the 16quant mode represents each empirical PDF by equal-probability quantile nodes and propagates their Cartesian products through the same projected-pair geometry. These modes differ in computational cost and approximation strategy, but are designed to estimate the same underlying PDF-marginalized pair-count target.

The LOS pair-response diagnostic provides a direct test of the pair-level redistribution induced by the PDFs. For DESI galaxy pairs selected at fixed true projected separation, the empirical-PDF, GMM, MC, and 16quant modes produce very similar smooth response matrices in estimated LOS separation. The point-estimate photo-$z$ case is more structured and more concentrated, as expected when the full radial distribution is reduced to a single scalar redshift. The corresponding cumulative inclusion curves show the basic trade-off that controls projected measurements: increasing the LOS window retains more physically close pairs but also admits more foreground--background leakage. 

A further diagnostic based on pair-probability calibration shows directly where the information gain from full PDFs enters. Full-PDF pair weights provide a continuous and nearly calibrated ranking of true close-pair membership, whereas point-photo-$z$ estimates make overconfident binary assignments.  For example, at
$\Pi_{\max}=400\,h^{-1}{\rm Mpc}$, point-photo-$z$ rejected pairs still
contain a true-close fraction of about $25\%$, while accepted pairs are
truly close only about $77\%$ of the time. Thus PDF-aware estimators
preserve partial pair-membership information that is discarded by scalar
photo-$z$ compression.

The controlled CosmoDC2 degradation tests show how pair redistribution propagates into the final projected clustering signal. In single-Gaussian and two-component surrogate experiments, the true clustering signal is progressively displaced from small to large $\pi$ as the imposed radial uncertainty increases. The PDF-aware cumulative $w_p(r_\mathrm{p};\Pi)$ curves therefore rise more slowly than the truth-redshift curve, but recover the projected truth signal once the LOS integration depth is large enough to include the redistributed close-pair probability. The recovery depth $\Pi^{\rm rec}_{\max}$ follows an approximately linear relation with the effective pairwise radial dispersion, with only weak dependence on projected separation over the range tested. This provides an important practical interpretation: the LOS integration scale is not arbitrary, but should be tied to the pairwise radial smearing of the sample.

The experiments with full empirical PDFs are closer to the regime expected in large photometric surveys. In the CosmoDC2 test, the full-PDF MC measurement is closer to the truth-redshift clustering than the point-estimate photo-$z$ measurement when a sufficiently broad projected window is used. However, this comparison also highlights a central limitation for real photometric surveys such as LSST by the Vera Rubin Observatory. In the mock, the truth-redshift cumulative curve is known and can be used to define a recovery depth. In real photometric data, this curve is an unknown quantity. Therefore, the integration depth must be calibrated indirectly, using the PDFs themselves, realistic simulations, spectroscopic validation samples, and stability tests of $w_p(r_\mathrm{p};\Pi)$ as the LOS window is enlarged. The cumulative curves remain valuable in this setting, but they diagnose convergence and leakage rather than providing an exact recovery point.

The parameter-convergence tests show that the approximate PDF modes reach stable results with practical choices of their internal resolution parameters. For the empirical-PDF CosmoDC2 subsample, MC measurements improve from $n_{\rm real}=10$ to $25$, with only modest additional gains at $n_{\rm real}=100$. The 16quant mode is already accurate for $q=8$ and improves slightly up to $q=16$--32, making $q\simeq16$ a useful default compromise. The GMM mode benefits strongly from increasing from $K=1$ to $K=2$, with smaller gains at $K=3$, reflecting the need for more than one component to represent asymmetric or multi-peaked empirical PDFs. Overall, MC and 16quant show the fastest practical convergence in the tests performed here, while GMM remains useful when a compact analytic LOS representation is desired. The reduced exact-grid benchmark presented in
the Appendix~\ref{app:exact_benchmark} confirms that, at the full estimator level, the approximate modes agree with the deterministic empirical-PDF result. It also illustrates the substantially greater computational cost of exact-grid integration, and the much higher performance of the Monte Carlo approach.

These results support the main methodological conclusion of this work: projected clustering with photometric redshift PDFs is best understood as a controlled redistribution problem. Full PDFs do not simply change the amplitude of $w_p(r_\mathrm{p})$; they move pair probability across the LOS and, in general, projected-separation bins. When this redistribution is propagated at the pair-count level and the LOS integration scale is chosen consistently with the pairwise radial uncertainty, the projected clustering signal can be recovered with small residual biases in the controlled tests. At the same time, the choice of $\Pi_{\max}$ becomes part of the measurement definition and should be reported, calibrated, and tested rather than treated as a fixed technical setting.

The methods developed here may also support applications beyond the validation experiments considered in this work. For photometric BAO analyses, $\nu$Gundam can provide PDF-marginalized pair counts for testing how realistic redshift uncertainties affect the transverse acoustic feature \citep{chan2022,ferreira2026}, although a complete analysis would still require survey-window modeling, covariance estimation, and BAO-template fitting. At smaller scales, the same machinery is directly applicable to 
probabilistic close-pair measurements, where the redshift PDFs of both 
galaxies are combined to estimate their probability of physical association \citep[e.g.][]{lopezsanjuan2015,mundy2017,duncan2019,dougherty2024,duan2025}. With extensions for physical apertures, sample selection, completeness corrections, and companion normalization, $\nu$Gundam could provide an efficient backend for merger-fraction studies in large photometric surveys.

A future extension of \scalebox{1.2}{$\nu$}\textsc{Gundam}\xspace will support cross-correlations between spectroscopic and PDF-based samples \citep{myers2009}. Anchoring one member of each pair with a precise redshift reduces the effective radial uncertainty, improving LOS localization and signal-to-noise while enabling the clustering of large photometric samples to be measured against sparse spectroscopic tracers within the same PDF-aware framework. Clustering redshifts allow to measure the positions of galaxies with unknown distances against reference objects with known locations to map cosmic structures \citep{Newman_2008, 2020A&A...642A.200V}. A direct application of \scalebox{1.2}{$\nu$}\textsc{Gundam}\xspace to improve photometric redshift estimations through clustering is currently under development for the Dark Energy Science Collaboration (DESC) photometric redshift data challenge.

\begin{acknowledgements}
      Part of this work was supported by \emph{ESO}, project
      number Ts~17/2--1.
      U.S. acknowledges support from the Polish National Science Center (NCN) grant UMO-2024/53/B/ST9/00230.
\end{acknowledgements}

\bibliographystyle{aa}
\bibliography{nug.bib}

@ARTICLE{smith2019,
       author = {{Smith}, Alex and {He}, Jian-hua and {Cole}, Shaun and {Stothert}, Lee and {Norberg}, Peder and {Baugh}, Carlton and {Bianchi}, Davide and {Wilson}, Michael J. and {Brooks}, David and {Forero-Romero}, Jaime E. and {Moustakas}, John and {Percival}, Will J. and {Tarl{\'e}}, Gregory and {Wechsler}, Risa H.},
        title = "{Correcting for fibre assignment incompleteness in the DESI Bright Galaxy Survey}",
      journal = {\mnras},
         year = 2019,
        month = mar,
       volume = {484},
       number = {1},
        pages = {1285-1300},
          doi = {10.1093/mnras/stz059},
archivePrefix = {arXiv},
       eprint = {1809.07355}
}

@ARTICLE{bianchi2025,
       author = {{Bianchi}, D. and {Hanif}, M. M. S. and {Carnero Rosell}, A. and {Lasker}, J. and {Ross}, A. J. and {Pinon}, M. and {de Mattia}, A. and {White}, M. and others},
        title = "{Characterization of DESI fiber assignment incompleteness effect on 2-point clustering and mitigation methods for DR1 analysis}",
      journal = {\jcap},
         year = 2025,
        month = apr,
       volume = {2025},
       number = {04},
          eid = {074},
        pages = {074},
          doi = {10.1088/1475-7516/2025/04/074},
archivePrefix = {arXiv},
       eprint = {2411.12025}
}

@ARTICLE{salvato2019,
  author  = {{Salvato}, M. and {Ilbert}, O. and {Hoyle}, B.},
  title   = {The many flavours of photometric redshifts},
  journal = {Nature Astronomy},
  year    = {2019},
  volume  = {3},
  pages   = {212--222},
  doi     = {10.1038/s41550-018-0478-0}
}

@ARTICLE{schmidt2020,
  author  = {{Schmidt}, S. J. and {Malz}, A. I. and {Soo}, J. Y. H. and others},
  title   = {Evaluation of probabilistic photometric redshift estimation approaches for The Rubin Observatory Legacy Survey of Space and Time (LSST)},
  journal = {MNRAS},
  year    = {2020},
  volume  = {499},
  number  = {2},
  pages   = {1587--1606},
  doi     = {10.1093/mnras/staa2799}
}

@ARTICLE{georgakakis2014,
  author  = {{Georgakakis}, A. and {Mountrichas}, G. and {Salvato}, M. and others},
  title   = {Large-scale clustering measurements with photometric redshifts: comparing the dark matter haloes of X-ray AGN, star-forming and passive galaxies at z $\approx$ 1},
  journal = {MNRAS},
  year    = {2014},
  volume  = {443},
  number  = {4},
  pages   = {3327--3340},
  doi     = {10.1093/mnras/stu1326}
}

@ARTICLE{allevato2016,
  author  = {{Allevato}, V. and {Marchesi}, S. and {Miyaji}, T. and others},
  title   = {The clustering of X-Ray-selected active galactic nuclei at z = 3},
  journal = {ApJ},
  year    = {2016},
  volume  = {832},
  number  = {1},
  eid     = {70},
  doi     = {10.3847/0004-637X/832/1/70}
}

@ARTICLE{lopezsanjuan2015,
  author  = {{L{\'o}pez-Sanjuan}, C. and {Cenarro}, A. J. and {Hern{\'a}ndez-Monteagudo}, C. and others},
  title   = {The ALHAMBRA survey: An empirical estimation of the cosmic variance for merger fraction studies based on close pairs},
  journal = {A\&A},
  year    = {2015},
  volume  = {576},
  eid     = {A53},
  doi     = {10.1051/0004-6361/201424913}
}

@ARTICLE{mundy2017,
  author  = {{Mundy}, C. J. and {Conselice}, C. J. and {Duncan}, K. J. and {Almaini}, O. and {H{\"a}u{\ss}ler}, B. and {Hartley}, W. G.},
  title   = {The evolution of the galaxy merger rate since z $\sim$ 3: identifying mergers using probabilistic pair counts},
  journal = {MNRAS},
  year    = {2017},
  volume  = {470},
  number  = {3},
  pages   = {3507--3531},
  doi     = {10.1093/mnras/stx1238}
}

@ARTICLE{duncan2019,
  author  = {{Duncan}, K. and {Conselice}, C. J. and {Mundy}, C. J. and others},
  title   = {Observational constraints on the merger history of galaxies since z $\approx$ 6: probabilistic galaxy pair counts in the UKIDSS Ultra Deep Survey},
  journal = {ApJ},
  year    = {2019},
  volume  = {876},
  number  = {2},
  eid     = {110},
  doi     = {10.3847/1538-4357/ab148a}
}

@ARTICLE{carrascokind2013,
  author  = {{Carrasco Kind}, M. and {Brunner}, R. J.},
  title   = {TPZ: photometric redshift PDFs and ancillary information by using prediction trees and random forests},
  journal = {MNRAS},
  year    = {2013},
  volume  = {432},
  number  = {2},
  pages   = {1483--1501},
  doi     = {10.1093/mnras/stt574}
}

@ARTICLE{wittman2016,
  author  = {{Wittman}, D. and {Bhaskar}, R. and {Tobin}, R.},
  title   = {Overconfidence in photometric redshift estimation},
  journal = {MNRAS},
  year    = {2016},
  volume  = {457},
  number  = {4},
  pages   = {4005--4011},
  doi     = {10.1093/mnras/stw261}
}

@ARTICLE{alonso2012,
  author        = {{Alonso}, D.},
  title         = {CUTE solutions for two-point correlation functions from large cosmological datasets},
  journal       = {arXiv e-prints},
  year          = {2012},
  eid           = {arXiv:1210.1833},
  eprint        = {1210.1833},
  archivePrefix = {arXiv},
  primaryClass  = {astro-ph.CO}
}

@ARTICLE{sinha2020,
  author  = {{Sinha}, M. and {Garrison}, L. H.},
  title   = {CORRFUNC: blazing fast correlation functions on the CPU},
  journal = {MNRAS},
  year    = {2020},
  volume  = {491},
  number  = {2},
  pages   = {3022--3041},
  doi     = {10.1093/mnras/stz3157}
}

@ARTICLE{hand2018,
  author  = {{Hand}, N. and {Feng}, Y. and {Beutler}, F. and others},
  title   = {nbodykit: an open-source, massively parallel toolkit for large-scale structure},
  journal = {AJ},
  year    = {2018},
  volume  = {156},
  number  = {4},
  eid     = {160},
  doi     = {10.3847/1538-3881/aadae0}
}

@ARTICLE{zhao2023,
  author  = {{Zhao}, C.},
  title   = {Fast correlation function calculator (FCFC): a high-performance pair-counting toolkit},
  journal = {A\&A},
  year    = {2023},
  volume  = {672},
  eid     = {A83},
  doi     = {10.1051/0004-6361/202346015}
}

@ARTICLE{norberg2009,
  author  = {{Norberg}, P. and {Baugh}, C. M. and {Gazta{\~n}aga}, E. and {Croton}, D. J.},
  title   = {Statistical analysis of galaxy surveys -- I. Robust error estimation for two-point clustering statistics},
  journal = {MNRAS},
  year    = {2009},
  volume  = {396},
  number  = {1},
  pages   = {19--38},
  doi     = {10.1111/j.1365-2966.2009.14389.x}
}

@ARTICLE{zehavi2011,
  author  = {{Zehavi}, I. and {Zheng}, Z. and {Weinberg}, D. H. and others},
  title   = {Galaxy clustering in the completed SDSS redshift survey: the dependence on color and luminosity},
  journal = {ApJ},
  year    = {2011},
  volume  = {736},
  number  = {1},
  eid     = {59},
  doi     = {10.1088/0004-637X/736/1/59}
}

@ARTICLE{peebleshauser1974,
  author  = {{Peebles}, P. J. E. and {Hauser}, M. G.},
  title   = {Statistical analysis of catalogs of extragalactic objects. III. The Shane-Wirtanen and Zwicky catalogs},
  journal = {ApJS},
  year    = {1974},
  volume  = {28},
  pages   = {19--36},
  doi     = {10.1086/190308}
}

@ARTICLE{malz2018,
  author  = {{Malz}, A. I. and {Marshall}, P. J. and {DeRose}, J. and {Graham}, M. L. and {Schmidt}, S. J. and {Wechsler}, R. H.},
  title   = {Approximating photo-z PDFs for large surveys},
  journal = {AJ},
  year    = {2018},
  volume  = {156},
  number  = {1},
  eid     = {35},
  doi     = {10.3847/1538-3881/aac6b5}
}

@ARTICLE{WISE,
       author = {{Wright}, Edward L. and {Eisenhardt}, Peter R.~M. and {Mainzer}, Amy K. and {Ressler}, Michael E. and {Cutri}, Roc M. and {Jarrett}, Thomas and {Kirkpatrick}, J. Davy and {Padgett}, Deborah and {McMillan}, Robert S. and {Skrutskie}, Michael and {Stanford}, S.~A. and {Cohen}, Martin and {Walker}, Russell G. and {Mather}, John C. and {Leisawitz}, David and {Gautier}, III, Thomas N. and {McLean}, Ian and {Benford}, Dominic and {Lonsdale}, Carol J. and {Blain}, Andrew and {Mendez}, Bryan and {Irace}, William R. and {Duval}, Valerie and {Liu}, Fengchuan and {Royer}, Don and {Heinrichsen}, Ingolf and {Howard}, Joan and {Shannon}, Mark and {Kendall}, Martha and {Walsh}, Amy L. and {Larsen}, Mark and {Cardon}, Joel G. and {Schick}, Scott and {Schwalm}, Mark and {Abid}, Mohamed and {Fabinsky}, Beth and {Naes}, Larry and {Tsai}, Chao-Wei},
        title = "{The Wide-field Infrared Survey Explorer (WISE): Mission Description and Initial On-orbit Performance}",
      journal = {\aj},
         year = 2010,
        month = dec,
       volume = {140},
       number = {6},
        pages = {1868-1881},
          doi = {10.1088/0004-6256/140/6/1868},
archivePrefix = {arXiv},
       eprint = {1008.0031},
 primaryClass = {astro-ph.IM},
       adsurl = {https://ui.adsabs.harvard.edu/abs/2010AJ....140.1868W}
}

@article{ferreira2026,
doi = {10.1088/1475-7516/2026/04/003},
url = {https://doi.org/10.1088/1475-7516/2026/04/003},
year = {2026},
month = {apr},
publisher = {IOP Publishing},
volume = {2026},
number = {04},
pages = {003},
author = {Ferreira, Paula S. and Reis, Ribamar R.R.},
title = {Influence of photometric galaxies redshift distribution in BAO estimation},
journal = {Journal of Cosmology and Astroparticle Physics}
}

@article{dougherty2024,
    author = {Dougherty, Sean L and Harrison, C M and Kocevski, Dale D and Rosario, D J},
    title = {Obscured AGN enhancement in galaxy pairs at cosmic noon: evidence from a probabilistic treatment of photometric redshifts},
    journal = {Monthly Notices of the Royal Astronomical Society},
    volume = {527},
    number = {2},
    pages = {3146-3163},
    year = {2024},
    month = {01},
    issn = {0035-8711},
    doi = {10.1093/mnras/stad1300},
    url = {https://doi.org/10.1093/mnras/stad1300},
    eprint = {https://academic.oup.com/mnras/article-pdf/527/2/3146/53663718/stad1300.pdf},
}

@ARTICLE{duan2025,
       author = {{Duan}, Qiao and {Conselice}, Christopher J. and {Li}, Qiong and {Austin}, Duncan and {Harvey}, Thomas and {Adams}, Nathan J. and {Duncan}, Kenneth J. and {Trussler}, James and {Ferreira}, Leonardo and {Westcott}, Lewi and {Harris}, Honor and {Windhorst}, Rogier A. and {Holwerda}, Benne W. and {Broadhurst}, Thomas J. and {Coe}, Dan and {Cohen}, Seth H. and {Du}, Xiaojing and {Driver}, Simon P. and {Frye}, Brenda and {Grogin}, Norman A. and {Hathi}, Nimish P. and {Jansen}, Rolf A. and {Koekemoer}, Anton M. and {Marshall}, Madeline A. and {Nonino}, Mario and {Ortiz}, III, Rafael and {Pirzkal}, Nor and {Robotham}, Aaron and {Ryan}, Russell E. and {Summers}, Jake and {D'Silva}, Jordan C.~J. and {Willmer}, Christopher N.~A. and {Yan}, Haojing},
        title = "{Galaxy mergers in the epoch of reionization ─ I. A JWST study of pair fractions, merger rates, and stellar mass accretion rates at z = 4.5─11.5}",
      journal = {\mnras},
         year = 2025,
        month = jun,
       volume = {540},
       number = {1},
        pages = {774-805},
          doi = {10.1093/mnras/staf638},
archivePrefix = {arXiv},
       eprint = {2407.09472},
 primaryClass = {astro-ph.GA},
       adsurl = {https://ui.adsabs.harvard.edu/abs/2025MNRAS.540..774D}
}

@article{RandomForest,
    author       = {{Breiman}, L.},
    title        = {{Random Forests}},
    journal      = {Machine Learning},
    year         = {2001},
    volume       = {45},
    pages        = {5--32},
    doi          = {10.1023/A:1010933404324},
}

@article{SKlearn_ML,
    author       = {{Pedregosa}, F. and {Varoquaux}, G. and {Gramfort}, A. and
                    others},
    title        = {{Scikit-learn: Machine Learning in {P}ython}},
    journal      = {Journal of Machine Learning Research},
    year         = {2011},
    volume       = {12},
    pages        = {2825--2830},
}

@ARTICLE{DESC_DC2,
       author = {{LSST Dark Energy Science Collaboration (LSST DESC)} and {Abolfathi}, Bela and {Alonso}, David and {Armstrong}, Robert and {Aubourg}, {\'E}ric and {Awan}, Humna and {Babuji}, Yadu N. and {Bauer}, Franz Erik and {Bean}, Rachel and {Beckett}, George and {Biswas}, Rahul and {Bogart}, Joanne R. and {Boutigny}, Dominique and {Chard}, Kyle and {Chiang}, James and {Claver}, Chuck F. and {Cohen-Tanugi}, Johann and {Combet}, C{\'e}line and {Connolly}, Andrew J. and {Daniel}, Scott F. and {Digel}, Seth W. and {Drlica-Wagner}, Alex and {Dubois}, Richard and {Gangler}, Emmanuel and {Gawiser}, Eric and {Glanzman}, Thomas and {Gris}, Phillipe and {Habib}, Salman and {Hearin}, Andrew P. and {Heitmann}, Katrin and {Hernandez}, Fabio and {Hlo{\v{z}}ek}, Ren{\'e}e and {Hollowed}, Joseph and {Ishak}, Mustapha and {Ivezi{\'c}}, {\v{Z}}eljko and {Jarvis}, Mike and {Jha}, Saurabh W. and {Kahn}, Steven M. and {Kalmbach}, J. Bryce and {Kelly}, Heather M. and {Kovacs}, Eve and {Korytov}, Danila and {Krughoff}, K. Simon and {Lage}, Craig S. and {Lanusse}, Fran{\c{c}}ois and {Larsen}, Patricia and {Le Guillou}, Laurent and {Li}, Nan and {Longley}, Emily Phillips and {Lupton}, Robert H. and {Mandelbaum}, Rachel and {Mao}, Yao-Yuan and {Marshall}, Phil and {Meyers}, Joshua E. and {Moniez}, Marc and {Morrison}, Christopher B. and {Nomerotski}, Andrei and {O'Connor}, Paul and {Park}, HyeYun and {Park}, Ji Won and {Peloton}, Julien and {Perrefort}, Daniel and {Perry}, James and {Plaszczynski}, St{\'e}phane and {Pope}, Adrian and {Rasmussen}, Andrew and {Reil}, Kevin and {Roodman}, Aaron J. and {Rykoff}, Eli S. and {S{\'a}nchez}, F. Javier and {Schmidt}, Samuel J. and {Scolnic}, Daniel and {Stubbs}, Christopher W. and {Tyson}, J. Anthony and {Uram}, Thomas D. and {Villarreal}, Antonia Sierra and {Walter}, Christopher W. and {Wiesner}, Matthew P. and {Wood-Vasey}, W. Michael and {Zuntz}, Joe},
        title = "{The LSST DESC DC2 Simulated Sky Survey}",
      journal = {\apjs},
         year = 2021,
        month = mar,
       volume = {253},
       number = {1},
          eid = {31},
        pages = {31},
          doi = {10.3847/1538-4365/abd62c},
archivePrefix = {arXiv},
       eprint = {2010.05926},
 primaryClass = {astro-ph.IM},
       adsurl = {https://ui.adsabs.harvard.edu/abs/2021ApJS..253...31L}
}

@article{zhou2021_desi,
    author = {Zhou, Rongpu and Newman, Jeffrey A and Mao, Yao-Yuan and Meisner, Aaron and Moustakas, John and Myers, Adam D and Prakash, Abhishek and Zentner, Andrew R and Brooks, David and Duan, Yutong and Landriau, Martin and Levi, Michael E and Prada, Francisco and Tarle, Gregory},
    title = {The clustering of DESI-like luminous red galaxies using photometric redshifts},
    journal = {Monthly Notices of the Royal Astronomical Society},
    volume = {501},
    number = {3},
    pages = {3309-3331},
    year = {2021},
    month = {03},
    issn = {0035-8711},
    doi = {10.1093/mnras/staa3764},
    url = {https://doi.org/10.1093/mnras/staa3764},
    eprint = {https://academic.oup.com/mnras/article-pdf/501/3/3309/35654181/staa3764.pdf},
}

@ARTICLE{DESI2022,
       author = {{DESI Collaboration} and {Abareshi}, B. and {Aguilar}, J. and {Ahlen}, S. and {Alam}, Shadab and {Alexander}, David M. and {Alfarsy}, R. and {Allen}, L. and {Allende Prieto}, C. and {Alves}, O. and {Ameel}, J. and {Armengaud}, E. and {Asorey}, J. and {Aviles}, Alejandro and {Bailey}, S. and {Balaguera-Antol{\'\i}nez}, A. and {Ballester}, O. and {Baltay}, C. and {Bault}, A. and {Beltran}, S.~F. and {Benavides}, B. and {BenZvi}, S. and {Berti}, A. and {Besuner}, R. and {Beutler}, Florian and {Bianchi}, D. and {Blake}, C. and {Blanc}, P. and {Blum}, R. and {Bolton}, A. and {Bose}, S. and {Bramall}, D. and {Brieden}, S. and {Brodzeller}, A. and {Brooks}, D. and {Brownewell}, C. and {Buckley-Geer}, E. and {Cahn}, R.~N. and {Cai}, Z. and {Canning}, R. and {Capasso}, R. and {Carnero Rosell}, A. and {Carton}, P. and {Casas}, R. and {Castander}, F.~J. and {Cervantes-Cota}, J.~L. and {Chabanier}, S. and {Chaussidon}, E. and {Chuang}, C. and {Circosta}, C. and {Cole}, S. and {Cooper}, A.~P. and {da Costa}, L. and {Cousinou}, M.-C. and {Cuceu}, A. and {Davis}, T.~M. and {Dawson}, K. and {de la Cruz-Noriega}, R. and {de la Macorra}, A. and {de Mattia}, A. and {Della Costa}, J. and {Demmer}, P. and {Derwent}, M. and {Dey}, A. and {Dey}, B. and {Dhungana}, G. and {Ding}, Z. and {Dobson}, C. and {Doel}, P. and {Donald-McCann}, J. and {Donaldson}, J. and {Douglass}, K. and {Duan}, Y. and {Dunlop}, P. and {Edelstein}, J. and {Eftekharzadeh}, S. and {Eisenstein}, D.~J. and {Enriquez-Vargas}, M. and {Escoffier}, S. and {Evatt}, M. and {Fagrelius}, P. and {Fan}, X. and {Fanning}, K. and {Fawcett}, V.~A. and {Ferraro}, S. and {Ereza}, J. and {Flaugher}, B. and {Font-Ribera}, A. and {Forero-Romero}, J.~E. and {Frenk}, C.~S. and {Fromenteau}, S. and {G{\"a}nsicke}, B.~T. and {Garcia-Quintero}, C. and {Garrison}, L. and {Gazta{\~n}aga}, E. and {Gerardi}, F. and {Gil-Mar{\'\i}n}, H. and {Gontcho A Gontcho}, S. and {Gonzalez-Morales}, Alma X. and {Gonzalez-de-Rivera}, G. and {Gonzalez-Perez}, V. and {Gordon}, C. and {Graur}, O. and {Green}, D. and {Grove}, C. and {Gruen}, D. and {Gutierrez}, G. and {Guy}, J. and {Hahn}, C. and {Harris}, S. and {Herrera}, D. and {Herrera-Alcantar}, Hiram K. and {Honscheid}, K. and {Howlett}, C. and {Huterer}, D. and {Ir{\v{s}}i{\v{c}}}, V. and {Ishak}, M. and {Jelinsky}, P. and {Jiang}, L. and {Jimenez}, J. and {Jing}, Y.~P. and {Joyce}, R. and {Jullo}, E. and {Juneau}, S. and {Kara{\c{c}}ayl{\i}}, N.~G. and {Karamanis}, M. and {Karcher}, A. and {Karim}, T. and {Kehoe}, R. and {Kent}, S. and {Kirkby}, D. and {Kisner}, T. and {Kitaura}, F. and {Koposov}, S.~E. and {Kov{\'a}cs}, A. and {Kremin}, A. and {Krolewski}, Alex and {L'Huillier}, B. and {Lahav}, O. and {Lambert}, A. and {Lamman}, C. and {Lan}, Ting-Wen and {Landriau}, M. and {Lane}, S. and {Lang}, D. and {Lange}, J.~U. and {Lasker}, J. and {Le Guillou}, L. and {Leauthaud}, A. and {Le Van Suu}, A. and {Levi}, Michael E. and {Li}, T.~S. and {Magneville}, C. and {Manera}, M. and {Manser}, Christopher J. and {Marshall}, B. and {Martini}, Paul and {McCollam}, W. and {McDonald}, P. and {Meisner}, Aaron M. and {Mena-Fern{\'a}ndez}, J. and {Meneses-Rizo}, J. and {Mezcua}, M. and {Miller}, T. and {Miquel}, R. and {Montero-Camacho}, P. and {Moon}, J. and {Moustakas}, J. and {Mueller}, E. and {Mu{\~n}oz-Guti{\'e}rrez}, Andrea and {Myers}, Adam D. and {Nadathur}, S. and {Najita}, J. and {Napolitano}, L. and {Neilsen}, E. and {Newman}, Jeffrey A. and {Nie}, J.~D. and {Ning}, Y. and {Niz}, G. and {Norberg}, P. and {Noriega}, Hern{\'a}n E. and {O'Brien}, T. and {Obuljen}, A. and {Palanque-Delabrouille}, N. and {Palmese}, A. and {Zhiwei}, P. and {Pappalardo}, D. and {PENG}, X. and {Percival}, W.~J. and {Perruchot}, S. and {Pogge}, R. and {Poppett}, C. and {Porredon}, A. and {Prada}, F. and {Prochaska}, J. and {Pucha}, R. and {P{\'e}rez-Fern{\'a}ndez}, A. and {P{\'e}rez-R{\`a}fols}, I. and {Rabinowitz}, D. and {Raichoor}, A.},
        title = "{Overview of the Instrumentation for the Dark Energy Spectroscopic Instrument}",
      journal = {\aj},
         year = 2022,
        month = nov,
       volume = {164},
       number = {5},
          eid = {207},
        pages = {207},
          doi = {10.3847/1538-3881/ac882b},
archivePrefix = {arXiv},
       eprint = {2205.10939},
 primaryClass = {astro-ph.IM},
       adsurl = {https://ui.adsabs.harvard.edu/abs/2022AJ....164..207D}
}

@ARTICLE{DESI_dr1_2025,
       author = {{DESI Collaboration} and {Karim}, M. Abdul and {Adame}, A.~G. and {Aguado}, D. and {Aguilar}, J. and {Ahlen}, S. and {Alam}, S. and {Aldering}, G. and {Alexander}, D.~M. and {Alfarsy}, R. and {Allen}, L. and {Allende Prieto}, C. and {Alves}, O. and {Anand}, A. and {Andrade}, U. and {Armengaud}, E. and {Avila}, S. and {Aviles}, A. and {Awan}, H. and {Bailey}, S. and {Baleato Lizancos}, A. and {Ballester}, O. and {Bault}, A. and {Bautista}, J. and {Bean}, R. and {Behera}, J. and {BenZvi}, S. and {Beraldo e Silva}, L. and {Bermejo-Climent}, J.~R. and {Beutler}, F. and {Bianchi}, D. and {Blake}, C. and {Blum}, R. and {Bolton}, A.~S. and {Bonici}, M. and {Brieden}, S. and {Brodzeller}, A. and {Brooks}, D. and {Buckley-Geer}, E. and {Burtin}, E. and {Bystr{\"o}m}, A. and {Canning}, R. and {Carnero Rosell}, A. and {Carr}, A. and {Carrilho}, P. and {Casas}, L. and {Castander}, F.~J. and {Cereskaite}, R. and {Cervantes-Cota}, J.~L. and {Chaussidon}, E. and {Chaves-Montero}, J. and {Chen}, S. and {Chen}, X. and {Circosta}, C. and {Claybaugh}, T. and {Cole}, S. and {Cooper}, A.~P. and {Cousinou}, M.-C. and {Cuceu}, A. and {Davis}, T.~M. and {Dawson}, K.~S. and {de Belsunce}, R. and {de la Cruz}, R. and {de la Macorra}, A. and {de Mattia}, A. and {Deiosso}, N. and {Della Costa}, J. and {Demina}, R. and {Demirbozan}, U. and {DeRose}, J. and {Dey}, A. and {Dey}, B. and {Ding}, J. and {Ding}, Z. and {Doel}, P. and {Douglass}, K. and {Dowicz}, M. and {Ebina}, H. and {Edelstein}, J. and {Eisenstein}, D.~J. and {Elbers}, W. and {Emas}, N. and {Escoffier}, S. and {Fagrelius}, P. and {Fan}, X. and {Fanning}, K. and {Favole}, G. and {Fawcett}, V.~A. and {Fern{\'a}ndez-Garc{\'\i}a}, E. and {Ferraro}, S. and {Findlay}, N. and {Font-Ribera}, A. and {Forero-Romero}, J.~E. and {Forero-S{\'a}nchez}, D. and {Frenk}, C.~S. and {G{\"a}nsicke}, B.~T. and {Galbany}, L. and {Garc{\'\i}a-Bellido}, J. and {Garcia-Quintero}, C. and {Garrison}, L.~H. and {Gazta{\~n}aga}, E. and {Gil-Mar{\'\i}n}, H. and {Gloudemans}, A. and {Gnedin}, O.~Y. and {Gontcho}, S. Gontcho A and {Gonzalez}, D. and {Gonzalez-Morales}, A.~X. and {Gonzalez-Perez}, V. and {Gordon}, C. and {Graur}, O. and {Green}, D. and {Gruen}, D. and {Gsponer}, R. and {Guandalin}, C. and {Gutierrez}, G. and {Guy}, J. and {Hahn}, C. and {Han}, J.~J. and {Han}, J. and {He}, S. and {Herrera-Alcantar}, H.~K. and {Heydenreich}, S. and {Honscheid}, K. and {Hou}, J. and {Howlett}, C. and {Huterer}, D. and {Ir{\v{s}}i{\v{c}}}, V. and {Ishak}, M. and {Jacques}, A. and {Jiang}, L. and {Jimenez}, J. and {Jing}, Y.~P. and {Joachimi}, B. and {Joudaki}, S. and {Joyce}, R. and {Jullo}, E. and {Juneau}, S. and {Kara{\c{c}}ayl{\i}}, N.~G. and {Karim}, T. and {Kehoe}, R. and {Kent}, S. and {Khederlarian}, A. and {Kirkby}, D. and {Kisner}, T. and {Kitaura}, F.-S. and {Kizhuprakkat}, N. and {Kong}, H. and {Koposov}, S.~E. and {Kremin}, A. and {Krolewski}, A. and {Lahav}, O. and {Lai}, Y. and {Lamman}, C. and {Lan}, T.-W. and {Landriau}, M. and {Lang}, D. and {Lange}, J.~U. and {Lasker}, J. and {Le Goff}, J.~M. and {Le Guillou}, L. and {Leauthaud}, A. and {Levi}, M.~E. and {Li}, S. and {Li}, T.~S. and {Liu}, W. and {Lodha}, K. and {Lokken}, M. and {Luo}, Y. and {Luo}, Y. and {Magneville}, C. and {Manera}, M. and {Manser}, C.~J. and {Margala}, D. and {Martini}, P. and {Maus}, M. and {McCullough}, J. and {McDonald}, P. and {Medina}, G.~E. and {Medina-Varela}, L. and {Meisner}, A. and {Mena-Fern{\'a}ndez}, J. and {Menegas}, A. and {Meneses-Rizo}, J. and {Mezcua}, M. and {Miquel}, R. and {Montero-Camacho}, P. and {Moon}, J. and {Moustakas}, J. and {Mu{\~n}oz-Guti{\'e}rrez}, A. and {Mu{\~n}oz-Santos}, D. and {Myers}, A.~D. and {Myles}, J. and {Nadathur}, S. and {Najita}, J. and {Napolitano}, L. and {Newman}, J.~A. and {Nikakhtar}, F. and {Nikutta}, R. and {Niz}, G. and {Noriega}, H.~E.},
        title = "{Data Release 1 of the Dark Energy Spectroscopic Instrument}",
      journal = {arXiv e-prints},
         year = 2025,
        month = mar,
          eid = {arXiv:2503.14745},
        pages = {arXiv:2503.14745},
          doi = {10.48550/arXiv.2503.14745},
archivePrefix = {arXiv},
       eprint = {2503.14745},
 primaryClass = {astro-ph.CO},
       adsurl = {https://ui.adsabs.harvard.edu/abs/2025arXiv250314745D}
}

@ARTICLE{donoso2019,
       author = {{Donoso}, E.},
        title = "{GUNDAM: a toolkit for fast spatial correlation functions in galaxy surveys}",
      journal = {\mnras},
         year = 2019,
        month = aug,
       volume = {487},
       number = {2},
        pages = {2824-2835},
          doi = {10.1093/mnras/stz1469},
       adsurl = {https://ui.adsabs.harvard.edu/abs/2019MNRAS.487.2824D}
}

@ARTICLE{myers2009,
       author = {{Myers}, Adam D. and {White}, Martin and {Ball}, Nicholas M.},
        title = "{Incorporating photometric redshift probability density information into real-space clustering measurements}",
      journal = {\mnras},
         year = 2009,
        month = nov,
       volume = {399},
       number = {4},
        pages = {2279-2287},
          doi = {10.1111/j.1365-2966.2009.15432.x},
archivePrefix = {arXiv},
       eprint = {0903.3121},
 primaryClass = {astro-ph.CO},
       adsurl = {https://ui.adsabs.harvard.edu/abs/2009MNRAS.399.2279M}
}

@ARTICLE{arnaltemur2009,
       author = {{Arnalte-Mur}, Pablo and {Fern{\'a}ndez-Soto}, Alberto and {Mart{\'\i}nez}, Vicent J. and {Saar}, Enn and {Hein{\"a}m{\"a}ki}, Pekka and {Suhhonenko}, Ivan},
        title = "{Recovering the real-space correlation function from photometric redshift surveys}",
      journal = {\mnras},
         year = 2009,
        month = apr,
       volume = {394},
       number = {3},
        pages = {1631-1639},
          doi = {10.1111/j.1365-2966.2009.14430.x},
archivePrefix = {arXiv},
       eprint = {0812.4226},
 primaryClass = {astro-ph},
       adsurl = {https://ui.adsabs.harvard.edu/abs/2009MNRAS.394.1631A}
}

@ARTICLE{asorey2016,
       author = {{Asorey}, J. and {Carrasco Kind}, M. and {Sevilla-Noarbe}, I. and {Brunner}, R.~J. and {Thaler}, J.},
        title = "{Galaxy clustering with photometric surveys using PDF redshift information}",
      journal = {\mnras},
         year = 2016,
        month = jun,
       volume = {459},
       number = {2},
        pages = {1293-1309},
          doi = {10.1093/mnras/stw721},
archivePrefix = {arXiv},
       eprint = {1601.00357},
 primaryClass = {astro-ph.CO},
       adsurl = {https://ui.adsabs.harvard.edu/abs/2016MNRAS.459.1293A}
}

@ARTICLE{chavesmontero2018,
       author = {{Chaves-Montero}, Jon{\'a}s and {Angulo}, Ra{\'u}l E. and {Hern{\'a}ndez-Monteagudo}, Carlos},
        title = "{The effect of photometric redshift uncertainties on galaxy clustering and baryonic acoustic oscillations}",
      journal = {\mnras},
         year = 2018,
        month = jul,
       volume = {477},
       number = {3},
        pages = {3892-3909},
          doi = {10.1093/mnras/sty924},
archivePrefix = {arXiv},
       eprint = {1610.09688},
 primaryClass = {astro-ph.CO},
       adsurl = {https://ui.adsabs.harvard.edu/abs/2018MNRAS.477.3892C}
}

@ARTICLE{chan2022,
       author = {{Chan}, Kwan Chuen and {Ferrero}, Ismael and {Avila}, Santiago and {Ross}, Ashley J. and {Crocce}, Martin and {Gazta{\~n}aga}, Enrique},
        title = "{Clustering with general photo-z uncertainties: application to Baryon Acoustic Oscillations}",
      journal = {\mnras},
         year = 2022,
        month = apr,
       volume = {511},
       number = {3},
        pages = {3965-3982},
          doi = {10.1093/mnras/stac340},
archivePrefix = {arXiv},
       eprint = {2110.13332},
 primaryClass = {astro-ph.CO},
       adsurl = {https://ui.adsabs.harvard.edu/abs/2022MNRAS.511.3965C}
}

@ARTICLE{DESI2016,
       author = {{DESI Collaboration} and {Aghamousa}, Amir and {Aguilar}, Jessica and {Ahlen}, Steve and {Alam}, Shadab and {Allen}, Lori E. and {Allende Prieto}, Carlos and {Annis}, James and {Bailey}, Stephen and {Balland}, Christophe and {Ballester}, Otger and {Baltay}, Charles and {Beaufore}, Lucas and {Bebek}, Chris and {Beers}, Timothy C. and {Bell}, Eric F. and {Bernal}, Jos{\'e} Luis and {Besuner}, Robert and {Beutler}, Florian and {Blake}, Chris and {Bleuler}, Hannes and {Blomqvist}, Michael and {Blum}, Robert and {Bolton}, Adam S. and {Briceno}, Cesar and {Brooks}, David and {Brownstein}, Joel R. and {Buckley-Geer}, Elizabeth and {Burden}, Angela and {Burtin}, Etienne and {Busca}, Nicolas G. and {Cahn}, Robert N. and {Cai}, Yan-Chuan and {Cardiel-Sas}, Laia and {Carlberg}, Raymond G. and {Carton}, Pierre-Henri and {Casas}, Ricard and {Castander}, Francisco J. and {Cervantes-Cota}, Jorge L. and {Claybaugh}, Todd M. and {Close}, Madeline and {Coker}, Carl T. and {Cole}, Shaun and {Comparat}, Johan and {Cooper}, Andrew P. and {Cousinou}, M.-C. and {Crocce}, Martin and {Cuby}, Jean-Gabriel and {Cunningham}, Daniel P. and {Davis}, Tamara M. and {Dawson}, Kyle S. and {de la Macorra}, Axel and {De Vicente}, Juan and {Delubac}, Timoth{\'e}e and {Derwent}, Mark and {Dey}, Arjun and {Dhungana}, Govinda and {Ding}, Zhejie and {Doel}, Peter and {Duan}, Yutong T. and {Ealet}, Anne and {Edelstein}, Jerry and {Eftekharzadeh}, Sarah and {Eisenstein}, Daniel J. and {Elliott}, Ann and {Escoffier}, St{\'e}phanie and {Evatt}, Matthew and {Fagrelius}, Parker and {Fan}, Xiaohui and {Fanning}, Kevin and {Farahi}, Arya and {Farihi}, Jay and {Favole}, Ginevra and {Feng}, Yu and {Fernandez}, Enrique and {Findlay}, Joseph R. and {Finkbeiner}, Douglas P. and {Fitzpatrick}, Michael J. and {Flaugher}, Brenna and {Flender}, Samuel and {Font-Ribera}, Andreu and {Forero-Romero}, Jaime E. and {Fosalba}, Pablo and {Frenk}, Carlos S. and {Fumagalli}, Michele and {Gaensicke}, Boris T. and {Gallo}, Giuseppe and {Garcia-Bellido}, Juan and {Gaztanaga}, Enrique and {Pietro Gentile Fusillo}, Nicola and {Gerard}, Terry and {Gershkovich}, Irena and {Giannantonio}, Tommaso and {Gillet}, Denis and {Gonzalez-de-Rivera}, Guillermo and {Gonzalez-Perez}, Violeta and {Gott}, Shelby and {Graur}, Or and {Gutierrez}, Gaston and {Guy}, Julien and {Habib}, Salman and {Heetderks}, Henry and {Heetderks}, Ian and {Heitmann}, Katrin and {Hellwing}, Wojciech A. and {Herrera}, David A. and {Ho}, Shirley and {Holland}, Stephen and {Honscheid}, Klaus and {Huff}, Eric and {Hutchinson}, Timothy A. and {Huterer}, Dragan and {Hwang}, Ho Seong and {Illa Laguna}, Joseph Maria and {Ishikawa}, Yuzo and {Jacobs}, Dianna and {Jeffrey}, Niall and {Jelinsky}, Patrick and {Jennings}, Elise and {Jiang}, Linhua and {Jimenez}, Jorge and {Johnson}, Jennifer and {Joyce}, Richard and {Jullo}, Eric and {Juneau}, St{\'e}phanie and {Kama}, Sami and {Karcher}, Armin and {Karkar}, Sonia and {Kehoe}, Robert and {Kennamer}, Noble and {Kent}, Stephen and {Kilbinger}, Martin and {Kim}, Alex G. and {Kirkby}, David and {Kisner}, Theodore and {Kitanidis}, Ellie and {Kneib}, Jean-Paul and {Koposov}, Sergey and {Kovacs}, Eve and {Koyama}, Kazuya and {Kremin}, Anthony and {Kron}, Richard and {Kronig}, Luzius and {Kueter-Young}, Andrea and {Lacey}, Cedric G. and {Lafever}, Robin and {Lahav}, Ofer and {Lambert}, Andrew and {Lampton}, Michael and {Landriau}, Martin and {Lang}, Dustin and {Lauer}, Tod R. and {Le Goff}, Jean-Marc and {Le Guillou}, Laurent and {Le Van Suu}, Auguste and {Lee}, Jae Hyeon and {Lee}, Su-Jeong and {Leitner}, Daniela and {Lesser}, Michael and {Levi}, Michael E. and {L'Huillier}, Benjamin and {Li}, Baojiu and {Liang}, Ming and {Lin}, Huan and {Linder}, Eric and {Loebman}, Sarah R. and {Luki{\'c}}, Zarija and {Ma}, Jun and {MacCrann}, Niall and {Magneville}, Christophe and {Makarem}, Laleh and {Manera}, Marc and {Manser}, Christopher J. and {Marshall}, Robert and {Martini}, Paul and {Massey}, Richard and {Matheson}, Thomas and {McCauley}, Jeremy and {McDonald}, Patrick and {McGreer}, Ian D. and {Meisner}, Aaron and {Metcalfe}, Nigel and {Miller}, Timothy N. and {Miquel}, Ramon and {Moustakas}, John and {Myers}, Adam and {Naik}, Milind and {Newman}, Jeffrey A. and {Nichol}, Robert C. and {Nicola}, Andrina and {Nicolati da Costa}, Luiz and {Nie}, Jundan and {Niz}, Gustavo and {Norberg}, Peder and {Nord}, Brian and {Norman}, Dara and {Nugent}, Peter and {O'Brien}, Thomas and {Oh}, Minji and {Olsen}, Knut A.~G.},
        title = "{The DESI Experiment Part I: Science,Targeting, and Survey Design}",
      journal = {arXiv e-prints},
         year = 2016,
        month = oct,
          eid = {arXiv:1611.00036},
        pages = {arXiv:1611.00036},
          doi = {10.48550/arXiv.1611.00036},
archivePrefix = {arXiv},
       eprint = {1611.00036},
 primaryClass = {astro-ph.IM},
       adsurl = {https://ui.adsabs.harvard.edu/abs/2016arXiv161100036D}
}

@ARTICLE{korytov2019,
       author = {{Korytov}, Danila and {Hearin}, Andrew and {Kovacs}, Eve and {Larsen}, Patricia and {Rangel}, Esteban and {Hollowed}, Joseph and {Benson}, Andrew J. and {Heitmann}, Katrin and {Mao}, Yao-Yuan and {Bahmanyar}, Anita and {Chang}, Chihway and {Campbell}, Duncan and {DeRose}, Joseph and {Finkel}, Hal and {Frontiere}, Nicholas and {Gawiser}, Eric and {Habib}, Salman and {Joachimi}, Benjamin and {Lanusse}, Fran{\c{c}}ois and {Li}, Nan and {Mandelbaum}, Rachel and {Morrison}, Christopher and {Newman}, Jeffrey A. and {Pope}, Adrian and {Rykoff}, Eli and {Simet}, Melanie and {To}, Chun-Hao and {Vikraman}, Vinu and {Wechsler}, Risa H. and {White}, Martin and {(The LSST Dark Energy Science Collaboration}},
        title = "{CosmoDC2: A Synthetic Sky Catalog for Dark Energy Science with LSST}",
      journal = {\apjs},
         year = 2019,
        month = dec,
       volume = {245},
       number = {2},
          eid = {26},
        pages = {26},
          doi = {10.3847/1538-4365/ab510c},
archivePrefix = {arXiv},
       eprint = {1907.06530},
 primaryClass = {astro-ph.CO},
       adsurl = {https://ui.adsabs.harvard.edu/abs/2019ApJS..245...26K}
}

@ARTICLE{keihanen2019,
       author = {{Keih{\"a}nen}, E. and {Kurki-Suonio}, H. and {Lindholm}, V. and {Viitanen}, A. and {Suur-Uski}, A.-S. and {Allevato}, V. and {Branchini}, E. and {Marulli}, F. and {Norberg}, P. and {Tavagnacco}, D. and {de la Torre}, S. and {Valiviita}, J. and {Viel}, M. and {Bel}, J. and {Frailis}, M. and {S{\'a}nchez}, A.~G.},
        title = "{Estimating the galaxy two-point correlation function using a split random catalog}",
      journal = {\aap},
         year = 2019,
        month = nov,
       volume = {631},
          eid = {A73},
        pages = {A73},
          doi = {10.1051/0004-6361/201935828},
archivePrefix = {arXiv},
       eprint = {1905.01133},
 primaryClass = {astro-ph.CO},
       adsurl = {https://ui.adsabs.harvard.edu/abs/2019A&A...631A..73K}
}

@ARTICLE{davispeebles1983,
       author = {{Davis}, M. and {Peebles}, P.~J.~E.},
        title = "{A survey of galaxy redshifts. V. The two-point position and velocity correlations.}",
      journal = {\apj},
         year = 1983,
        month = apr,
       volume = {267},
        pages = {465-482},
          doi = {10.1086/160884},
       adsurl = {https://ui.adsabs.harvard.edu/abs/1983ApJ...267..465D}
}

@ARTICLE{sureshkumar2021,
       author = {{Sureshkumar}, U. and {Durkalec}, A. and {Pollo}, A. and {Bilicki}, M. and {Loveday}, J. and {Farrow}, D.~J. and {Holwerda}, B.~W. and {Hopkins}, A.~M. and {Liske}, J. and {Pimbblet}, K.~A. and {Taylor}, E.~N. and {Wright}, A.~H.},
        title = "{Galaxy and Mass Assembly (GAMA). Tracing galaxy environment using the marked correlation function}",
      journal = {\aap},
         year = 2021,
        month = sep,
       volume = {653},
          eid = {A35},
        pages = {A35},
          doi = {10.1051/0004-6361/202140500},
archivePrefix = {arXiv},
       eprint = {2102.04177},
 primaryClass = {astro-ph.CO},
       adsurl = {https://ui.adsabs.harvard.edu/abs/2021A&A...653A..35S}
}

@ARTICLE{skibba2006,
       author = {{Skibba}, Ramin and {Sheth}, Ravi K. and {Connolly}, Andrew J. and
        {Scranton}, Ryan},
        title = "{The luminosity-weighted or `marked' correlation function}",
      journal = {\mnras},
         year = 2006,
        month = Jun,
       volume = {369},
        pages = {68-76},
          doi = {10.1111/j.1365-2966.2006.10196.x},
archivePrefix = {arXiv},
       eprint = {astro-ph/0512463},
 primaryClass = {astro-ph},
       adsurl = {https://ui.adsabs.harvard.edu/#abs/2006MNRAS.369...68S}
}

@ARTICLE{gorski2005,
       author = {{G{\'o}rski}, K.~M. and {Hivon}, E. and {Banday}, A.~J. and {Wandelt}, B.~D. and {Hansen}, F.~K. and {Reinecke}, M. and {Bartelmann}, M.},
        title = "{HEALPix: A Framework for High-Resolution Discretization and Fast Analysis of Data Distributed on the Sphere}",
      journal = {\apj},
         year = 2005,
        month = apr,
       volume = {622},
        pages = {759-771},
          doi = {10.1086/427976},
       adsurl = {https://ui.adsabs.harvard.edu/abs/2005ApJ...622..759G}
}

@ARTICLE{landyszalay1993,
       author = {{Landy}, Stephen D. and {Szalay}, Alexander S.},
        title = "{Bias and Variance of Angular Correlation Functions}",
      journal = {\apj},
         year = 1993,
        month = jul,
       volume = {412},
        pages = {64},
          doi = {10.1086/172900},
       adsurl = {https://ui.adsabs.harvard.edu/abs/1993ApJ...412...64L}
}

@article{Newman_2008,
doi = {10.1086/589982},
url = {https://doi.org/10.1086/589982},
year = {2008},
month = {sep},
publisher = {},
volume = {684},
number = {1},
pages = {88},
author = {Newman, Jeffrey A.},
title = {Calibrating Redshift Distributions beyond Spectroscopic Limits with Cross-Correlations},
journal = {The Astrophysical Journal}
}

@ARTICLE{2020A&A...642A.200V,
       author = {{van den Busch}, J.~L. and {Hildebrandt}, H. and {Wright}, A.~H. and {Morrison}, C.~B. and {Blake}, C. and {Joachimi}, B. and {Erben}, T. and {Heymans}, C. and {Kuijken}, K. and {Taylor}, E.~N.},
        title = "{Testing KiDS cross-correlation redshifts with simulations}",
      journal = {\aap},
         year = 2020,
        month = oct,
       volume = {642},
          eid = {A200},
        pages = {A200},
          doi = {10.1051/0004-6361/202038835},
archivePrefix = {arXiv},
       eprint = {2007.01846},
 primaryClass = {astro-ph.CO},
       adsurl = {https://ui.adsabs.harvard.edu/abs/2020A&A...642A.200V}
}

\begin{appendix}
\nolinenumbers

\section{Data samples}
\label{app:data_samples}

\subsection{DESI}
\label{sec:DESI}

We use spectroscopic galaxy redshifts from the first public data release of the Dark Energy Spectroscopic Instrument \citep{DESI_dr1_2025}. DESI is a dark energy experiment mounted on the 4-meter Mayall telescope at Kitt Peak National Observatory, initially designed to measure spectroscopic redshifts for approximately 40 million galaxies and quasars over a footprint of 14\,000\,deg$^2$ 
(\citealt{DESI2016}, \citealt{DESI2022}).
The spectroscopic data were accessed from the \texttt{iron} spectroscopic production (the primary reduction pipeline associated with DR1 \citep{DESI_dr1_2025}, via the merged \textsc{HEALPix}-level redshift catalog \texttt{zall-pix-iron.fits}, which consolidates all targets across surveys and  programs into a single file organized by \textsc{HEALPix} pixel on the sky \citep{gorski2005}. Since projected clustering measurements require a reasonably uniform angular distribution of objects on the sky, we restricted the sample to a contiguous rectangular region defined by $128.9\degr \leq \mathrm{RA} \leq 228.1\degr$ and $-6.1\degr \leq \mathrm{Dec} \leq 3.9\degr$, covering an approximate area of 394\,deg$^2$. Retaining the full sample, including objects with potentially unreliable redshifts, is intentional, as one of the goals of this work is to assess the behavior of the PDF-aware estimators across a large range of photometric redshift quality. This sample contains 911\,491 galaxies.

\subsection{CosmoDC2}
\label{sec:cosmodc2}

CosmoDC2 is a large synthetic sky catalog developed for analysis of data from the Vera C. Rubin Observatory Legacy Survey of Space and Time \citep{korytov2019}. The catalog is built upon the Outer Rim N-body simulation and populated with galaxies using a combination of semi-analytic modeling and empirical abundance-matching techniques, providing realistic distributions of galaxy positions, shapes, magnitudes, and photometric properties across multiple broadband filters consistent with the LSST filter set ($ugrizy$). CosmoDC2 is made accessible through the Rubin Science Platform as part of Data Preview~0.2 \citep{DESC_DC2}, which provides simulated images and catalogs processed with the LSST Science Pipelines.

Objects were selected from the \texttt{dp02\_dc2\_catalogs.Object} table via the Table Access Protocol (TAP) service, retaining only primary detections (\texttt{detect\_isPrimary}\,=\,1) classified as galaxies (\texttt{truth\_type}\,=\,1 in the \texttt{TruthSummary} table), with magnitudes (computed from the \texttt{r\_cModelFlux} column via \texttt{scisql\_nanojanskyToAbMag}). Truth redshifts were obtained by joining the \texttt{Object} table with the \texttt{TruthSummary} table through the \texttt{MatchesTruth} table.

To ensure a uniform angular selection suitable for projected clustering measurements, we applied a depth-based mask using the survey limiting magnitude property map\footnote{\texttt{deepCoadd\_psf\_maglim\_consolidated\_map\_weighted\_mean}} provided with DP0.2, which encodes the weighted-mean point spread function (PSF) $5\sigma$ magnitude limit per \textsc{HEALPix} pixel. Only regions with a simulated depth greater than 26.7\,mag in the $r$ band were retained. An additional cut of $\mathrm{Dec} < -30\degr$ was applied to exclude the irregular upper boundary of the DC2 footprint. The resulting sample covers an approximate area of 211\,deg$^2$.

\subsection{Photometric redshift distributions}
\label{sec:photoz}

Photometric redshift probability density functions (PDFs) for both samples are estimated using independent Random Forest regressors trained on broadband photometry, implemented via the \texttt{scikit-learn} library \citep{SKlearn_ML}. Random Forests are ensemble methods that aggregate the predictions of a large number of decorrelated decision trees, each trained on a bootstrap resample of the training set and a random subset of the input features \citep{RandomForest}. In regression mode, each tree produces an independent scalar redshift prediction for a given object. Rather than retaining only the mean or median of these tree-level outputs, we exploit the full distribution of per-tree predictions to construct a non-parametric per-object redshift PDF. Specifically, for each galaxy the $T$ individual tree predictions $\{z_t^{(i)},\, t=1,\ldots,T\}$ are collected and used to populate a redshift histogram on a common grid, which after normalization yields the empirical PDF $p_i(z)$. This approach requires no assumption about the functional form of the redshift uncertainty and naturally captures asymmetries and multimodality. The photometric point estimates used in the point-redshift mode correspond to the median of the distribution.

Constructing redshift distributions from ensembles of prediction trees is related in spirit to the TPZ methodology \citep{carrascokind2013}. We treat the resulting histograms as empirical per-object uncertainty distributions rather than assuming that they are perfectly calibrated posteriors. As with other probabilistic photo-$z$ products, their widths and low-probability tails may be miscalibrated, and their adequacy should therefore be assessed with calibration and science-specific diagnostics \citep{wittman2016,schmidt2020}.

For the DESI sample, the input features consist of the optical magnitudes $g$, $r$, and $z$ from the DESI Legacy Imaging Surveys, together with the two mid-infrared bands $W1$ and $W2$ from the Wide-field Infrared Survey Explorer \citep{WISE}, as well as the colors between adjacent bands. The Random Forest was trained on all spectroscopic objects lying outside the rectangular footprint used for the clustering analysis (Sect.~\ref{sec:DESI}). This spatial separation was adopted for two reasons. First, removing objects from the clustering region to use them as training data would reduce the size of the analysis sample. Second, and more importantly, doing so could alter the real clustering signal present in that region, since the removed objects are physically correlated with their neighbors. The full clustering sample therefore constitutes an independent test set, on which the root mean square error (RMSE) of the photometric redshift 
point estimates is $\mathrm{RMSE} \approx 0.17$.

For the CosmoDC2 sample, the input features consist of the six broadband 
magnitudes ($ugrizy$) together with the five adjacent colors ($u-g$, $g-r$, 
$r-i$, $i-z$, $z-y$), computed from the \texttt{cModelFlux} measurements. 
Since truth redshifts are available for all objects, the Random Forest was 
trained on a randomly selected subset of five million galaxies drawn from the 
full catalog, with the remainder serving as the clustering analysis sample and 
test set (Sect.~\ref{sec:cosmodc2}). Since the training objects are selected 
randomly across the full catalog rather than by sky region, their removal does 
not affect the two-point statistics of the remaining sample: a random subsample 
preserves the clustering properties of the parent population. This strategy 
differs from the one adopted for DESI, where the relatively small size of the 
analysis region made it preferable to train on the spatially disjoint complement 
of that region, avoiding any reduction of the clustering sample. The RMSE 
evaluated on the test set is $\mathrm{RMSE} \approx 0.12$.

\section{Computational benchmark and exact-grid validation on a reduced sample}
\label{app:exact_benchmark}
The exact empirical-PDF mode provides the deterministic reference for the
PDF-marginalized pair-count calculation described in Sect.~\ref{sec:methods}. Its computational cost is nevertheless substantially higher than that of the compressed and stochastic approximations, particularly for broad empirical
PDFs, large line-of-sight integration ranges, and random catalogs several
times larger than the data sample. Repeating the complete controlled-degradation experiment of Sect.~\ref{subsec:validation:degradation} with the exact-grid mode would therefore be impractical. We instead perform a reduced-sample benchmark whose purpose is to test whether the production-oriented PDF modes reproduce the full estimator-level result obtained from exact empirical-PDF integration.

The benchmark uses the same CosmoDC2 population $0.1<z_{\rm true}<0.7$ and empirical photometric redshift PDFs considered in Sect.~\ref{subsec:validation:degradation}, among which we selected a random subsample of $\sim$200,000 data and $\sim$600,000 random galaxies. The calculation uses 12 projected-separation bins and 20 line-of-sight bins of width $\Delta\pi=20\,h^{-1}\,{\rm Mpc}$, corresponding to
$\Pi_{\max}\approx400\,h^{-1}\,{\rm Mpc}$, i.e. close to the $z_{\rm true}$ recovery scale. No bootstrap or jackknife resampling is included, so that the reported wall times isolate the cost of catalog preparation and the DD, DR, and RR pair counts.

For each approximate method $m$, we quantify agreement with the exact-grid
measurement through the fractional residual $\delta_m(r_\mathrm{p})\equiv(w_{p,m}(r_\mathrm{p})/{w_{p,{\rm ePDF}}(r_\mathrm{p}))-1}$, and summarize this residual over the projected-separation range using its root-mean-square value,
\begin{equation}
    {\rm RMS}_m
    =
    \left[
    \frac{1}{N_{r_\mathrm{p}}}
    \sum_b \delta_m^2(r_{\mathrm{p},b})
    \right]^{1/2},
    \label{eq:appendix_rms}
\end{equation}
and its median, which measures the typical signed fractional offset. Because
all modes are evaluated on precisely the same data and random catalogs, sample variance and random-catalog fluctuations are largely common to the compared curves.

Table~\ref{tab:exact_grid_benchmark} presents the resulting estimator-level
comparison. The exact-grid calculation uses
\texttt{edge\_refine=2} (see Sec. \ref{subsec:methods:exact-grid}) and
\texttt{prob\_floor=3.E-5}. The approximate calculations use
the practical configurations motivated by Sect.~\ref{sec:param_convergence}: $K=2$ Gaussian components for GMM, $N_q=16$ quantile nodes for 16quant, and $N_{\rm real}=25$ realizations for MC. The calculations were performed with 15 OpenMP threads on an Intel Core i9-9960X processor. The three approximate modes reproduce the exact-grid projected measurement with fractional RMS residuals below $5\%$. The closest overall agreement is obtained with 16quant, for which the RMS residual is $1.3\%$ and the median offset is $+1.4\%$. The MC result has a
larger RMS residual of $3.6\%$, as expected from the finite number of radial realizations, but its median offset is only $+0.1\%$. The GMM result has an RMS residual of $4.6\%$ and a median offset of $+4.2\%$. Its larger positive offset is consistent with the approximation discussed in Sect.~\ref{subsec:methods:gmm}, in which the LOS probability is integrated analytically but the projected separation is assigned from the component-pair means.

Relative to exact-grid integration, the measured speed-up factors are approximately $8.7$ for GMM, $28$ for 16quant, and $98$ for MC. The precise values depend on the PDF-grid resolution, probability floor, edge-refinement factor, LOS range, random-to-data ratio, and available
hardware. Nevertheless, the benchmark supports a clear practical
interpretation: exact-grid integration is valuable as a deterministic
validation reference for samples of small to moderate size, while GMM, 16quant, Monte Carlo
provide scalable alternatives for large samples.

\begin{table}
    \centering
    \caption{Reduced-sample benchmark of the PDF-aware projected-correlation modes.}
    \label{tab:exact_grid_benchmark}
    \begin{tabular}{lccc}
        \hline\hline
        Method &
        ${\rm RMS}\!\left( \delta_m \right)$ &
        ${\rm median}\!\left( \delta_m \right)$ &
        Wall time [min]\\
        \hline
        Exact ePDF &
        $-$ &
        $-$ &
        273 \\

        GMM &
        0.046 &
        +0.042 &
        31.4 (8.7x) \\

        16quant &
        0.013 &
        +0.014 &
        9.8 (28x)\\

        Monte Carlo &
        0.036 &
        +0.001 &
        2.8 (98x)\\
        \hline
    \end{tabular}
    \tablefoot{Fractional residuals are measured relative to the exact empirical-PDF result over the same $r_\mathrm{p}$ bins. All calculations use identical catalogs, PDF assignments, estimator settings, and $(r_\mathrm{p},\pi)$ binning. The numbers in parenthesis measure the speed-up factor respect to the baseline exact case.}
\end{table}

\FloatBarrier 
\clearpage

\end{appendix}
\end{document}